\documentclass[onecolumn,tightenlines,nofootinbib,preprintnumbers,
amsmath,amssymb]{revtex4}

\usepackage[usenames,dvipsnames]{color}
\usepackage{caption2}
\usepackage{dcolumn}
\usepackage{graphicx}
\usepackage{subfigure}
\usepackage{epstopdf}
\usepackage{bm}
\usepackage{lscape}
\renewcommand{\arraystretch}{0.9}
\usepackage{setspace}

\begin{document}
\begin{spacing}{1.5}

\title{Study of $K\pi$ final states from four-body deacy of  $\bar{B}_{s}$ meason  under perturbation QCD} 

\author{Jiao Wu $^{1}$,
	Na Wang $^{1}$,
	Gang L\"{u}$^{1}$\footnote{Email: ganglv66@sina.com},
	Xin-Heng Guo $^{2}$}

\affiliation{\small $^{1}$College of Physics, Henan University of Technology, Zhengzhou 450001, China\\
	$^{2}$ College of Nuclear Science and Technology, Beijing Normal University, Beijing 100875, China \\
}

\begin{abstract}	 
	 In the perturbative QCD framework, we use the quasi-two-body method to study the CP violations and branching ratios in the  $\bar B_{s} \rightarrow [\phi(\rho^{0},\omega)\rightarrow K^{+}K^{-}][K^{*0}\rightarrow K^{+}\pi^{-}]$  and $\bar B_{s} \rightarrow [\rho^{0}(\omega,\phi)\rightarrow (\pi^{+}\pi^{-})][K^{*0}\rightarrow K^{+}\pi^{-}]$ decay processes. Owing to the interference effects of $\rho^{0}$, $\omega$, and $\phi$, new strong phases associated with vector mesons will be generated, resulting in relatively large CP violation within the interference region. Moreover, we provide numerical comparisons of CP violation from resonance effects and non-resonance contributions. In order to provide better theoretical predictions for future experiments, we integrate CP violation over invariant mass to obtain the regional CP violation value. Furthermore, we have calculated the polarization fractions and branching ratios for different intermediate states in the four-body decay process, which may be influenced by vector interference effects. Additionally, we discuss the possibility of observing the predicted CP violation at the LHC.	  
\end{abstract}

\maketitle

\section{Introduction}

CP violation is a topic of great interest in the field of particle physics. Both theoretical and experimental efforts have been continuously exploring and searching for the origin of CP violation. In the Standard Model (SM), CP violation is mainly attributed to the Cabibbo-Kobayashi-Maskawa (CKM) matrix, which arises from the mixing and phase differences between quarks leading to CP violation in weak interaction \cite{Cabibbo:1963yz}. At the same time, some resonance effects are related to complex strong phares, which may affect the CP violation in the decay process \cite{Zhang:2013oqa}. The $\rho^{0}-\omega-\phi$ mixing mechanism can induce significant phase differences, leading to substantial CP violation within the interference range \cite{Li:2019xwh,Shi:2022ggo,Lu:2022rdi}. Since the discovery of CP violation in the  $K^{0} \bar{K}^{0} $ system in 1964 \cite{Christenson:1964fg},  the study of CP violation has become increasingly profound and precise due to continuous advancements in experimental techniques and an increase in data \cite{KTeV:1999aiu,Belle:2001zzw,BaBar:2001oxa}. In particular, the significant CP violation phenomenon observed during the multi-body decays of the B meson system provides a crucial experimental basis for in-depth research on the mechanism of CP violation\cite{LHCb:2012kja,LHCb:2012uja,Belle:2017cxf,LHCb:2023exl}. Therefore, multi-body decays have become a hot topic in recent years, providing clues for the revelation of new physics phenomena \cite{Bediaga:2013ela,Cheng:2016shb,Klein:2017xti,Li:2018lbd,Lu:2018fqe,Ma:2019qlm,LHCb:2019xmb,LHCb:2019jta,Yan:2022kck,Liang:2022mrz}.

 B mesons contain heavy b quarks, making reliable calculation of perturbation effects for their decay processes. The large branching ratios of B meson also provide excellent plate  for detecting CP violation \cite{Bander:1979px}. The four-body decay of B mesons is more complex compared to two-body and three-body decays due to complex phase space. Specifically, four-body decay requires consideration of interactions and kinematics among more final-state particles, as well as more decay pathways.  Due to the relatively limited research on B meson four-body decays, experimental and theoretical studies on this topic are of significant importance. Through in-depth investigation of the four-body decay process of B mesons, we can gain a more comprehensive understanding of the CP violation phenomena.  
 Unexpectedly, we observe that the interference of vector mesons induces significant alterations in the polarization fractions and branching ratios within certain four-body decay processes. This finding also offers a valuable theoretical reference for experimental studies on the polarization fractions and branching ratios of four-body decays.

In this work, we used the perturbative QCD (PQCD) method to calculate the CP violation and branching ratio
in the four-body decay process \cite{Lu:2000em,Keum:2000ph,Keum:2000wi}. The PQCD method has been successfully applied to study the non-leptonic two-body decay processes of B mesons, achieving significant  progress  \cite{Xiao:2006mg}. Typically,  the multibody decay process  involves sophisticated kinematics mechanisms, and the  calculation of the corresponding hadronic matrix elements is not simple.
  However, in the quasi-two-body approach, the four-body decay process can be viewed as a quasi-two-body decay process with an intermediate resonance state \cite{Zhang:2025jlt,Zou:2020fax,Hua:2020usv}.   Therefore, the  PQCD factorization formula for the studied four-body decay amplitude can be expressed as:
\setlength\abovedisplayskip{0.4cm}
\setlength\belowdisplayskip{0.4cm}
\begin{eqnarray}
		\begin{array}{ll}
	\mathcal{A} = A_{1}(\bar B_{s} \rightarrow V_1V_2) A_{2}(V_1\rightarrow  P_1 P_2) A_{3}(V_2\rightarrow  P_3 P_4),\\[7pt] 
	         		\end{array}     	
\end{eqnarray} 
for the decay process $\bar B_{s} \rightarrow V_1V_2$, its amplitude $A_{1}$ can be expressed as the convolution of the wave functions $\Phi_{B}$, $\Phi_{V_1} $, $\Phi_{V_2} $ and the  hard kernel function ${\cal H}$, that is $\Phi_{B} \otimes {\cal H} \otimes \Phi_{V_1} \otimes \Phi_{V_2}$.  Based on the formalism of Lepage, Brodsky, Botts, and Sterman, the PQCD factorization theorem has been developed for non-leptonic heavy meson decays \cite{Chang:1996dw,Yeh:1997rq,Lepage:1980fj,Botts:1989kf}. $A_{2}$ and $A_{3}$ represent the amplitudes of the $V_1\rightarrow  P_1 P_2$ and $V_2\rightarrow  P_3 P_4$ processes, respectively. 
Currently,  extensive research has been carried out on the two-body decays of B mesons, and a large amount of data and research results related to B meson decays have been accumulated,  including  the CP violation phenomenon and the branching  ratios \cite{Ali:2007ff}.
 Therefore, under a mixing mechanism, we utilized the quasi-two-body method to calculate the CP violation and branching ratio in the $\bar B_{s} \rightarrow {(K^{+} K^{-})( K^{+}\pi^{-})} $ and $\bar B_{s} \rightarrow {(\pi^{+} \pi^{-}) (K^{+}\pi^{-})} $ decay processes. By utilizing PQCD and quasi-two-body methods, we simplified the complex kinematics correlations, providing an effective computational framework for studying multi-body decays. 

 In the $\bar B_{s} \rightarrow V_1(V_1\rightarrow P_1 P_2)V_2(V_2 \rightarrow P_3 P_4)$ decay process, $V_1$ and $V_2$ act as intermediate vector particles which subsequently decay into two hadrons. Specifically, $V_1$ decays into $P_1 P_2$, while $V_2$ decays into $P_3 P_4$. By using the factorization relation, also known as the narrow width approximation (NWA), we can effectively decompose this decay process into a continuous two-body decay:
 \setlength\abovedisplayskip{0.4cm}
 \setlength\belowdisplayskip{0.4cm}
\begin{eqnarray}
	\Gamma\left(\bar B_{s} \rightarrow V_1(V_1\rightarrow P_1 P_2)V_2(V_2 \rightarrow P_3 P_4)\right)   =
	\Gamma\left(B \rightarrow V_1V_2\right) \mathcal{B}\left(V_1 \rightarrow  P_1 P_2\right)\mathcal{B}\left(V_2 \rightarrow P_3 P_4\right),
\end{eqnarray}
where $\Gamma$ represents the decay width and $\mathcal{B}$ represents the decay branching ratio. This method is only effective with narrow width limits, and it requires correction when the width becomes sufficiently large. Therefore, we introduce a correction factor:
 \setlength\abovedisplayskip{0.5cm}
\setlength\belowdisplayskip{0.5cm}
\begin{eqnarray}
	\eta _{v_{12}}=\frac{\Gamma \left( B\rightarrow V_1V_2 \right) \mathcal{B} \left( V_1\rightarrow P_1 P_2 \right) \mathcal{B} \left( V_2\rightarrow  P_3 P_4 \right)}{\Gamma \left( \bar B_{s} \rightarrow V_1(V_1\rightarrow P_1 P_2)V_2(V_2 \rightarrow P_3 P_4) \right)}.
\end{eqnarray}
The corrections to vector resonance typically fall below 10$\%$.  When applying the QCD factorization method, the correction factor for decay processes is approximately around 7$\%$ \cite{Cheng:2020iwk,Cheng:2020mna}.
When calculating CP violation, the constant $\eta_{v_{12}}$ can be eliminated without affecting the result,  thereby disregarding the impact of  the NWA on CP violation.

The article is structured as follows: In Sec. II, we provide a detailed explanation of the physical mechanism and calculation of mixing parameters based on the vector meson dominance (VMD) model, specifically focusing on the $\rho^{0}-\omega-\phi$ resonance effects. In Sec. III, we analyze the kinematics of the four-body decay process. In Sec. IV, we present typical Feynman diagrams and amplitude forms for first-order four-body decays within the framework of perturbative QCD (PQCD).  In Sec. V, we conduct a comprehensive analysis of CP violation in the $\bar B_{s} \rightarrow {(K^{+} K^{-}) (K^{+}\pi^{-})}$ and $\bar B_{s} \rightarrow {(\pi^{+} \pi^{-}) (K^{+}\pi^{-})}$ decay processes under the $\rho^{0}-\omega-\phi$ vector meson resonance mechanism. Relevant parameters are provided in Sec. VI. The numerical results for CP violation, regional CP violation, polarization fractions, and branching ratios are presented in Sec. VII. Finally, in Sec. VIII, we summarize our findings and discuss the feasibility of observing the predicted CP violation at the LHC.

\section{VECTOR MESON MIXING MECHANISM }

The vector meson dominance model (VMD) treats vector mesons as propagators interacting with photons \cite{Nambu:1957wzj,Kroll:1967it}. This model effectively elucidates the interaction between photons and hadrons. It plays a crucial role in our understanding of resonant states. Based on the VMD model, positron pairs annihilate to form photons. Subsequently, the resulting photons polarize in vacuum and generate vector particles such as $\rho^0(770)$, $\omega(782)$, and $\phi (1020)$. These vector particles then decay into  $\pi^{+}\pi^{-}$ meson pair \cite{Ivanov:1981wf,Achasov:2016lbc}. In VMD model, the electromagnetic form factor of the $\pi$ meson is needed to obtain the mixing parameters corresponding to the two vector particles \cite{OConnell:1995nse}.  The contribution of the vector meson dominance model to the $e^{+}e^{-}\rightarrow\pi^{+}\pi^{-}$ process is illustrated in Fig. 1 and Fig. 2.

\begin{figure}[h]
	\centering
	\includegraphics[height=3.3cm,width=8.05cm]{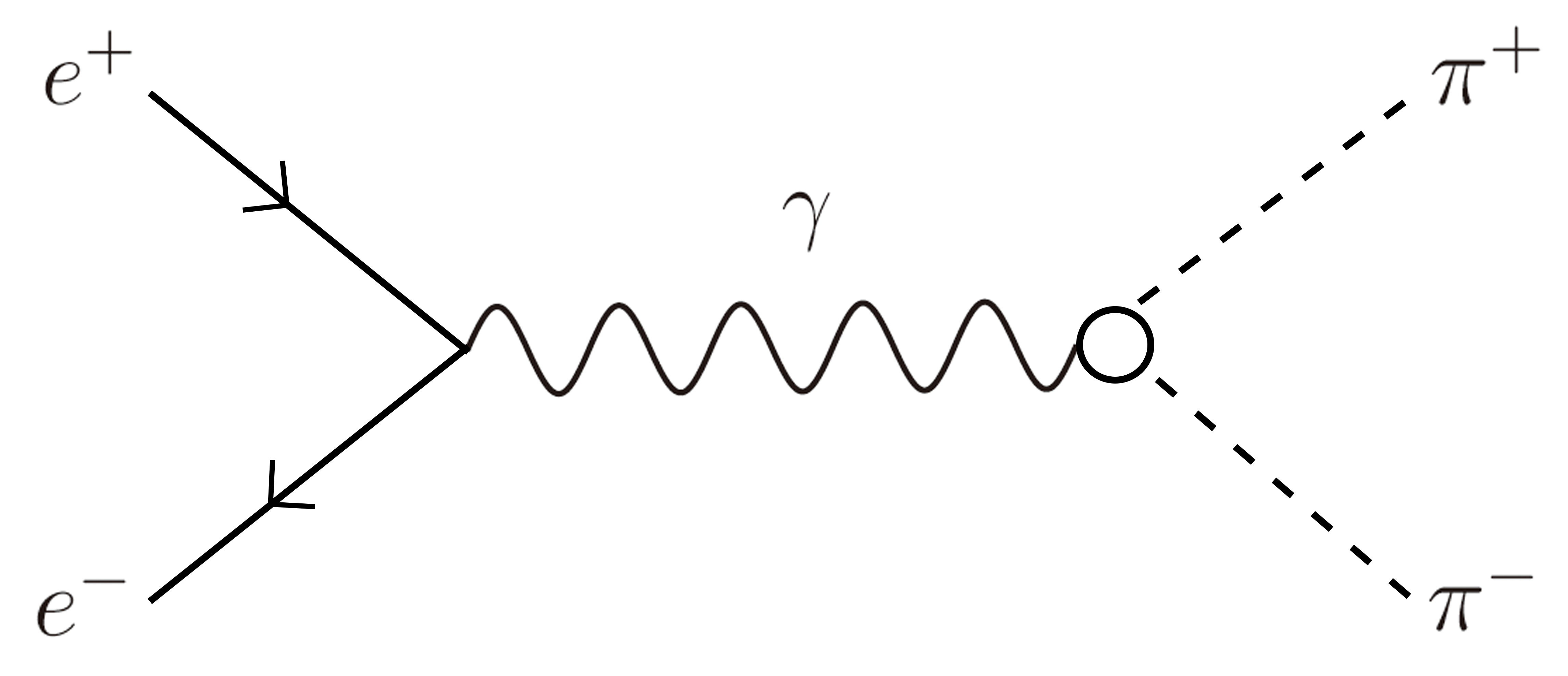}
	\caption{ In the S-channel, electron-positron pairs annihilate into a photon, which subsequently decays into a $\pi^+\pi^-$ pair.}
	\label{fig1}
\end{figure}

\begin{figure}[h]
	\centering
	\includegraphics[height=3.1cm,width=10.1cm]{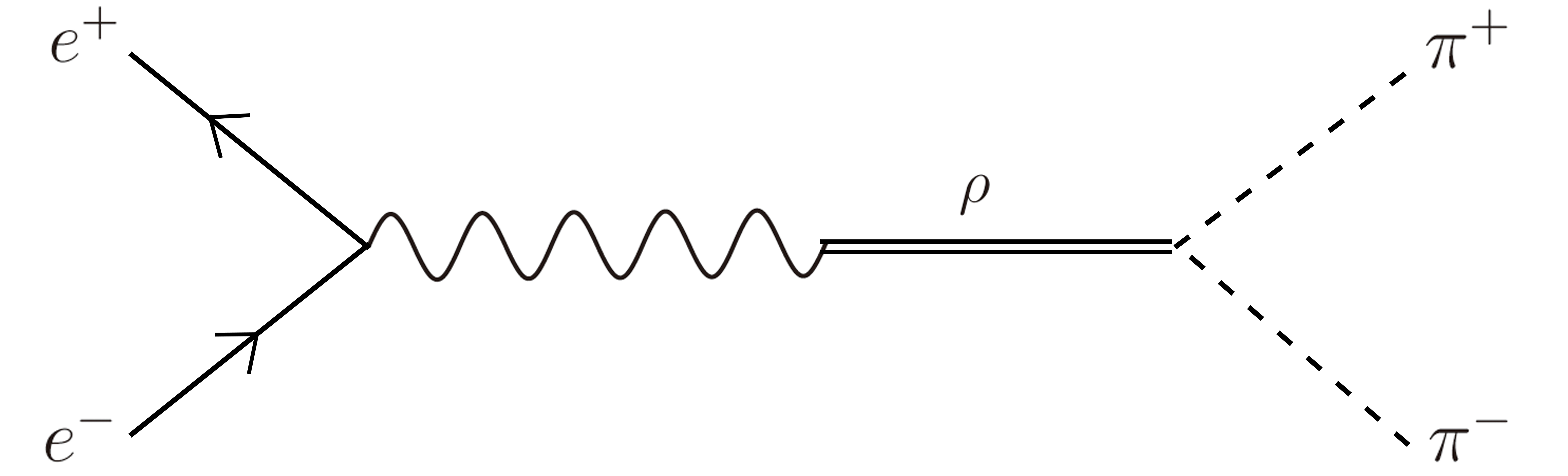}
	\caption{ In the S-channel, the Vector Meson Dominance (VMD) model is employed to describe the process of $e^{+}e^{-}\rightarrow\pi^{+}\pi^{-}$.}
	\label{fig2}
\end{figure}

The circle in Fig. 1 represents the form factor and represents all possible strong interactions that occur within the circle. We employ the VMD model to derive the description of the $e^{+}e^{-}\rightarrow\pi^{+}\pi^{-}$ process, as illustrated in Fig. 2.
The mechanism of $\rho-\omega$ mixing arises from quark mass differences and electromagnetic interaction effects. In the QCD Lagrangian, quark mass differences lead to isospin symmetry breaking, causing hadrons such as $\rho$ and $\omega$ to interact with photons, thereby inducing mixing. A similar mechanism applies to $\rho-\phi$ mixing. 
Therefore, the contribution of isospin breaking to the process $e^{+}e^{-}\rightarrow\pi^{+}\pi^{-}$ from $\rho-\omega$ mixing and $\rho-\phi$ mixing at leading order is illustrated in Fig. 3.

\begin{figure}[h]
	\centering
	\includegraphics[height=2.738cm,width=17cm]{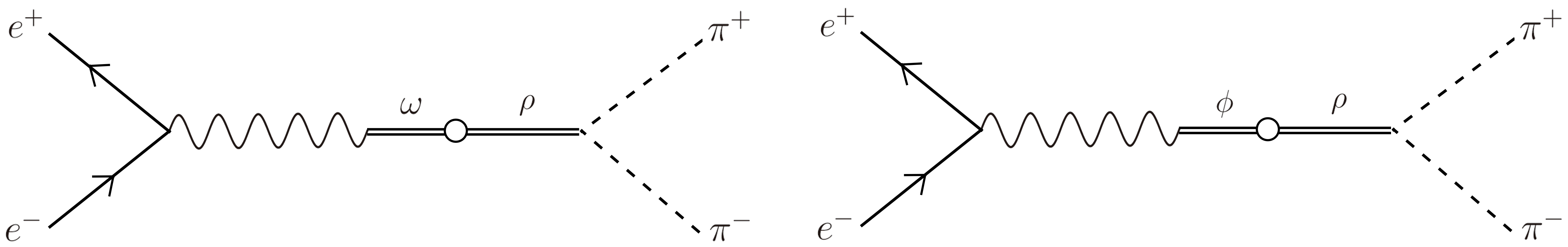}
	\caption{The contributions of $\rho-\omega$ and $\rho-\phi$ mixing to the process $e^{+}e^{-}\rightarrow\pi^{+}\pi^{-}$.}
	\label{fig3}
\end{figure}

Since the resonance state $ \rho^{0}- \omega-\phi$ belongs to a non-physical state, in order to obtain its expression in the physical field, we need to transform the isospin field into the physical field through the unitary matrix R(s) \cite{Lu:2022rdi}.
 There is a linear relationship between the physical state $ \rho^{0}- \omega-\phi$ and the isospin state $ \rho^{0}_{I}- \omega_{I}-\phi_{I}$, which is connected through the unitary matrix R(s): $A^{3\times1}=R(s)A^{3\times1}_{I}$. The matrix A comprises the elements $\rho^{0}$, $\omega$ and $\phi$, the matrix $A_{I}$ comprises the elements $\rho^{0}_{I}$, $\omega_{I}$ and $\phi_{I}$. And expression of R(s) is given as follows:
   \setlength\abovedisplayskip{0.5cm}
\setlength\belowdisplayskip{0.5cm}
\begin{equation}
	R(s)=
	\left (
	\begin{array}{lll}
		<\rho_{I}^0|\rho^0> & \hspace{0.3cm} <\omega_{I}|\rho^0>  &\hspace{0.3cm}<\phi_{I}|\rho^0>\\[0.5cm]
		<\rho_{I}^0|\omega> &  \hspace{0.3cm}<\omega_{I}|\omega>&\hspace{0.3cm}<\phi_{I}|\omega>\\[0.5cm]
		<\rho_{I}^0|\phi>&\hspace{0.3cm} <\omega_{I}|\phi> & \hspace{0.3cm} <\phi_{I}|\phi>
	\end{array}
	\right )
	\label{L1}	=\left(\begin{array}{ccc}
		1 &  \hspace{0.3cm}-F_{\rho^0 \omega}(s) &\hspace{0.3cm} -F_{\rho^0 \phi}(s) \\
		\\
		F_{\rho^0 \omega}(s) &\hspace{0.3cm} 1 &\hspace{0.3cm} -F_{\omega \phi}(s) \\
		\\
		F_{\rho^0 \phi}(s) &\hspace{0.3cm} F_{\omega \phi}(s) &\hspace{0.3cm} 1
	\end{array}\right),
\end{equation}
where $<\rho_{I}^0|\omega>$, $<\rho_{I}^0|\phi>$, $<\omega_{I}|\phi>$ and $F_{\rho^0\omega}(s)$, $F_{\rho^0\phi}(s)$, $F_{\omega\phi}(s)$ are all first-order approximations terms.
 Based on the isospin representation, we construct the isospin base vector  $|I,I_{3}>$, where $I$ and $I_3$ represent the isospin and its third component, respectively. 
 Subsequently, we used an orthogonal normalization relation to establish the relationship between physical states and isospin states. Then we can obtain the expression of ${\rho^0}$, $\omega$ and $\phi$ in the physical field:
   \setlength\abovedisplayskip{0.5cm}
\setlength\belowdisplayskip{0.5cm}
\setlength{\jot}{1pt}
\begin{equation}
	\begin{split}
		\rho^{0}=\rho_{I}^{0}-F_{\rho^0\omega }(s) \omega_{I}-F_{\rho^0\phi }(s) \phi_{I},\\ \\
		\omega=F_{ \rho^0\omega }(s) \rho_{I}^{0}+\omega_{I} -F_{\omega \phi}(s) \phi_{I}, \\ \\
		\phi=F_{\rho^0\phi  }(s) \rho_{I}^{0}+F_{\omega \phi}(s) \omega_{I}+\phi_{I}.
	\end{split}
\end{equation}

Furthermore, the mixed parameters $\Pi_{\rho^0\omega}$, $\Pi_{\rho^0\phi}$, $\Pi_{\omega\phi}$ and $F_{\rho^0\omega}$, $F_{\rho^0\phi}$, $F_{\omega\phi}$ are are order of $\mathcal{O}(\lambda)$ ($\lambda\ll 1$), while their multiplication yields higher order terms that can be disregarded in this process\cite{Lu:2022rdi}. Consequently, we obtain the expression of $F_{\rho^0\omega}$, $F_{\rho^0\phi}$, $F_{\omega\phi}$ as follows:
  \setlength\abovedisplayskip{0.6cm}
\setlength\belowdisplayskip{0.6cm}
\begin{equation}
	\begin{split}
		F_{\rho^0 \omega}=\frac{\Pi_{\rho^0 \omega}}{s_{\rho^0}-s_{\omega}},\hspace{0.3cm}\,  
		F_{\rho^0 \phi}=\frac{\Pi_{\rho^0 \phi}}{s_{\rho^0}-s_{\phi}},\hspace{0.3cm}\, 
		F_{\omega \phi}=\frac{\Pi_{\omega \phi}}{s_{\omega}-s_{\phi}}.
	\end{split}
\end{equation}

The relationship between $F_{V_1V_2}$=$-F_{V_2V_1}$ can be observed. We define the propagator by combining the meson width dependent on energy with the meson mass and momentum. $s_V$ represents the inverse propagator of the vector meson $(V=\rho^0$, $\omega$ or $\phi)$, and we define  $s_{V}=s-m_{V}^{2}+\mathrm{i} m_{V} \Gamma_{V}$. $m_V$ signifies the mass of the vector meson, $\Gamma_{V}$  is the decay width of a vector meson related to energy, and  $\sqrt{s}$ corresponds to the invariant mass of a pair of $P_1P_2$ mesons.

In this paper, we introduce the momentum-related mixed parameter $\Pi_{V_{1}V_{2}}$ to achieve a noticeable s-dependence \cite{Lu:2023yxa}. Wolfe and Maltnan accurately measured the mixed parameter $\Pi_{\rho^0 \omega }=-4470 \pm 250 \pm 160-i(5800 \pm 2000 \pm 1100)  \mathrm{MeV}^{2}$ near meson  $\rho^0$ \cite{Wolfe:2009ts,Wolfe:2010gf}. The mixing parameter  $\Pi_{\omega \phi}=19000+i(2500 \pm 300) \mathrm{MeV}^{2}$ is determined near the $\phi$ meson. Moreover, the mixed parameter  $\Pi_{\phi\rho^0}=720 \pm 180 -i(870 \pm 320)\mathrm{MeV}^{2}$ is obtained near the $\phi$ meson \cite{Achasov:1999wr}. Then we define :
\setlength\abovedisplayskip{0.5cm}
\setlength\belowdisplayskip{0.7cm}
\begin{equation} 
	\begin{split}
		\widetilde{\Pi}_{\rho^0\omega}=\frac{s_{\rho^0}\Pi_{\rho^0\omega}}{s_{\rho^0}-s_{\omega}},\hspace{0.3cm}\, ~~\widetilde{\Pi}_{\rho^0\phi}=\frac{s_{\rho^0}\Pi_{\rho^0\phi}}{s_{\rho^0}-s_{\phi}} ,\hspace{0.3cm}\, 
		~~\widetilde{\Pi}_{\phi\omega}=\frac{s_{\phi}\Pi_{\phi\omega}}{s_{\phi}-s_{\omega}} .
	\end{split}
\end{equation}

\section{FOUR-BODY DECAY KINEMATICS}

   Compared to the kinematics of two-body decay, the kinematics of four-body decay is significantly more intricate.  Currently, there is extensive research on the kinematics of four-body hadronic decay
   \cite{Pais:1968zza,Kane:1978ie,Lee:1992ih,Kramer:1992ag}.  The process of four-body decay not only encompasses contributions from resonant and non-resonant components but also involves interactions among the final-state particles \cite{Grozin:1983tt,Grozin:1986at,Muller:1994ses,Diehl:1998dk,Diehl:2000uv,Hagler:2002nf}.     
  The $\bar B_{s} \rightarrow V_1(V_1\rightarrow P_1 P_2)V_2(V_2 \rightarrow P_3 P_4)  $ process involves the decay of a $\bar B_{s}$ meson into two vector mesons, $V_1$ and $V_2$. Subsequently, $V_1$ decays into a pair of pseudoscalar mesons, $P_1$ and $P_2$, while $V_2$ decays into another pair of pseudoscalar mesons, $P_3$ and $P_4$.    
    \setlength\abovedisplayskip{0.8cm}
   \setlength\belowdisplayskip{0.6cm}
   \begin{figure}[h]
   	\centering
   	\includegraphics[height=5.495cm,width=11.025cm]{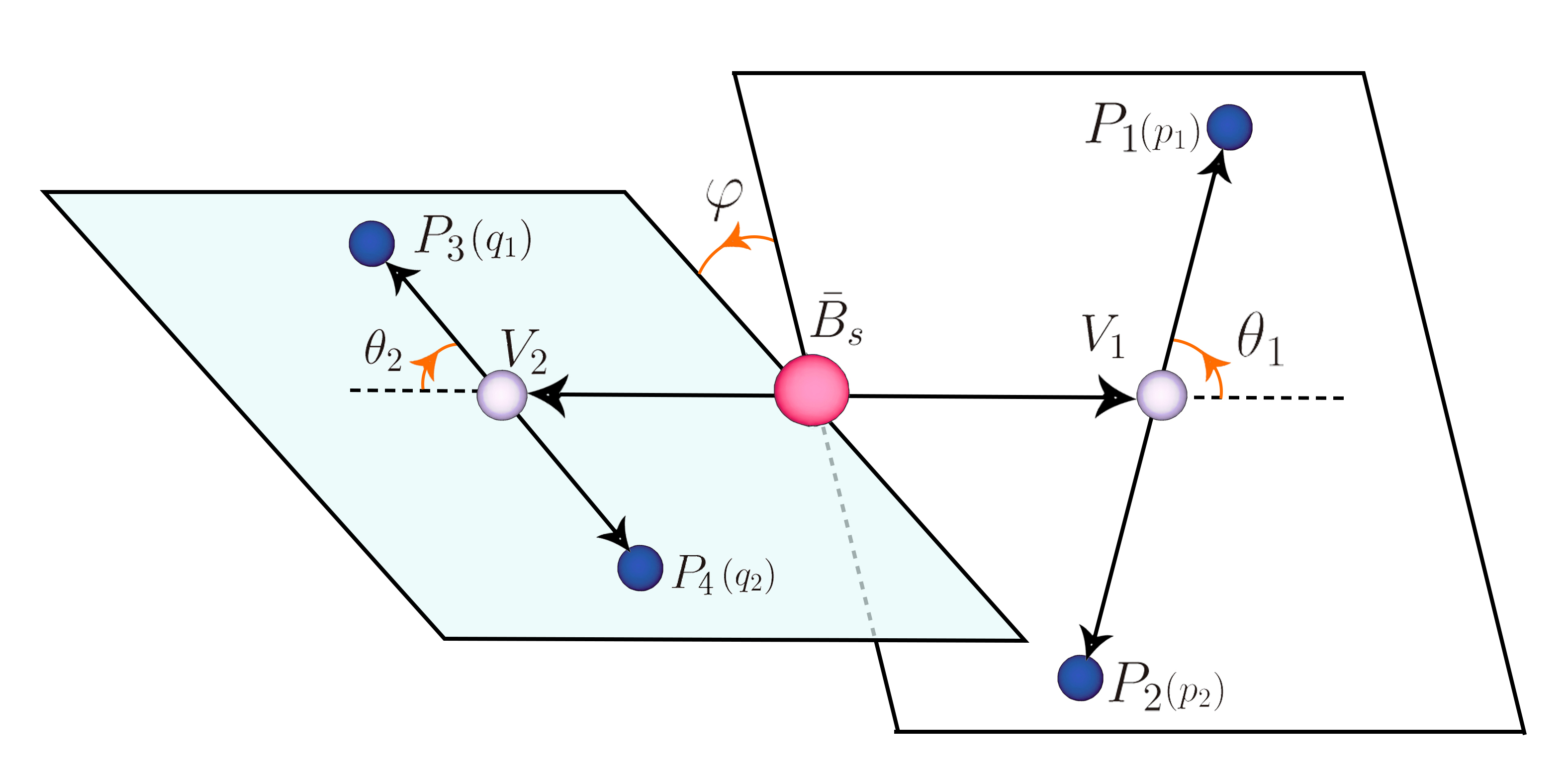}
   	\caption{ The decay diagrams of  $\bar B_{s} \rightarrow V_1(V_1\rightarrow P_1 P_2)V_2(V_2 \rightarrow P_3 P_4)$ process.}
   	\label{fig4}
   \end{figure}
   
    As illustrated in Fig. 4,  $P$, $p_{1}$, $p_{2}$, $q_{1}$ and $q_{2}$ denote the momenta of the $\bar B_{s}$ meson and the  final-state four mesons respectively.
    For $\bar B_{s}$ meson with initial spin 0, the phase space integral in its four-body decay process depends on five variables: the three helicity angles and the square of the invariant masses of the two final meson pairs.  These variables encompass:
     	
(1). \( s_1 = (p_1 + p_2)^2 \), the square of the invariant mass of the \( P_1 P_2 \) meson pair. 
 	
(2). \( s_2 = (q_1 + q_2)^2 \), the square of the invariant mass of the \( P_3 P_4 \) meson pair. 
 	
(3). \( \theta_1 \), the angle between the direction of motion of \( P_1 \) in the rest frame of the \( V_1 \) meson and the direction of motion of the \( V_1 \) meson in the rest frame of the \( \bar{B}_s \) meson.
 	
(4). \( \theta_2 \), the angle between the direction of motion of \( P_3 \) in the rest frame of the \( V_2 \) meson and the direction of motion of the \( V_2 \) meson in the rest frame of the \( \bar{B}_s \) meson.
 	 
(5). \( \varphi \), the angle between the plane defined by the \( P_1 P_2 \) meson pair and the plane defined by the \( P_3 P_4 \) meson pair in the rest frame of the \( \bar{B}_s \) meson.  
  
 $\Omega $ represents the phase space of four-body decay
  process,  where  $\mathrm{d} \Omega = X \alpha \beta ds_1 ds_2 d\cos \theta_1 d\cos \theta_2 d\varphi ~$. The partial decay width is defined as follows:   
  \setlength\abovedisplayskip{0.5cm}
  \setlength\belowdisplayskip{0.5cm}
\begin{eqnarray}
d\Gamma=\frac{|{\cal M}|^2}{4(4\pi)^6 m_B^3} X \alpha \beta ds_1 ds_2 d\cos \theta_1 d\cos \theta_2 d\varphi ~,
\end{eqnarray}
where $|{\cal M}|^2$  represents the square of the amplitude \cite{Hsiao:2017nga}. $X$, $\alpha$, and $\beta$ are represented respectively:
\begin{eqnarray}
 &&X=[(m_B^2-s_1-s_2)^2/4-s_1s_2]^{1/2}\,,\nonumber \\   
&&\alpha =\lambda^{1/2}(s_1,m_{p1}^2,m^{ 2}_{p2})/s_1\,,\nonumber\\  
&&\beta =\lambda^{1/2}(s_2,m_{q1}^2,m^{2}_{q2})/s_2\,.
\end{eqnarray}
with $\lambda(a,b,c)=a^2+b^2+c^2-2ab-2bc-2ca$. 

The permissible range of the five variables, on which the phase space integral depends, is defined as follows:
 \setlength\abovedisplayskip{0.3cm}
\setlength\belowdisplayskip{0.3cm}
\begin{eqnarray}
	(m_{p1}+m_{p2})^2\leq &\,s_1\,&\leq (m_{\bar B_s}-\sqrt{s_2})^2\,,\nonumber\\ 
	(m_{q1}+m_{q2})^2\leq &\,s_2\,&\leq (m_{\bar B_s}-m_{p1}-m_{p2})^2\,,\nonumber\\ 
	0\leq \theta_{1,2}\leq \pi&\,,&0\leq \varphi \leq 2\pi\,.
\end{eqnarray}
We can derive the branching ratio formula:
 \setlength\abovedisplayskip{0.3cm}
\setlength\belowdisplayskip{0.5cm}
\begin{eqnarray}
	\mathcal{B}=\frac{\tau_{\bar B_s}}{4(4\pi)^6 m_{\bar B_{s}}^3} \int{X \alpha \beta ds_1 ds_2 d\cos \theta_1 d\cos \theta_2 d\varphi|{\cal M}|^2} ~,
\end{eqnarray}
where $\tau_{\bar B_s}$ refers to the lifetime of $\bar B_s$ meson.

\section{THE AMPLITUDE OF THE QUASI-TWO-BODY DECAY PROCESS IN THE PERTURBATIVE QCD FRAME}

In the PQCD factorization method, the non-leptonic decay of B mesons is dominated by the exchange of hard gluons. The hard part of the decay process is separated and treated using perturbation theory, while the non-perturbative part is absorbed into the universal hadron wave function. According to the ``color transparency mechanism" in the initial state of the B-meson, the decay of the B meson causes the light quark to move at a high speed. For the spectator quark inside the B meson to gain large momentum and produce a fast-moving final meson with the light quark, a hard gluon is needed to provide energy.  Thus the two ``final state'' vector mesons continue to decay, eventually producing four final state particles.  The specific Feynman diagram is as follows:

\begin{figure}[h]
	\centering
	\includegraphics[height=6.46cm,width=17.17cm]{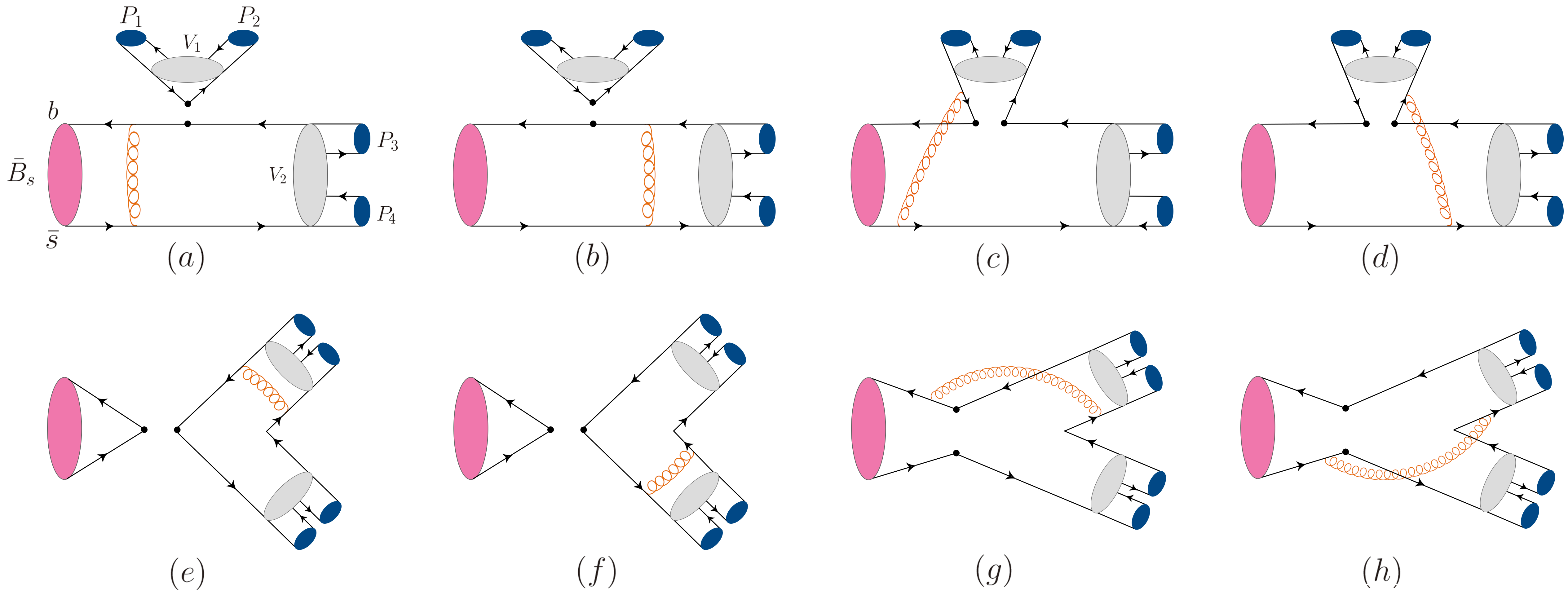}
	\caption{ The Feynman diagrams of $\bar B_{s} \rightarrow V_1(V_1\rightarrow P_1 P_2)V_2(V_2 \rightarrow P_3 P_4)  $ process.}
	\label{fig3}
\end{figure}

Typical Fig.5 is leading-order Feynman diagram for the four-body decay, where the symbol    \textbullet\medspace represents the weak interaction vertex. Diagrams (a)-(d) represents the emission contribution and diagrams (e)-(h) represents the annihilation contribution, with possible four-quark operator insertions. Using the quasi-two-body method, the total amplitude of  $\bar B_{s} \rightarrow V_1(V_1\rightarrow P_1 P_2)V_2(V_2 \rightarrow P_3 P_4)   $ consists of two components:  $\bar  B_{s} \rightarrow V_1V_2$  and $V_1V_2\rightarrow{(P_1 P_2 )(P_3 P_4)} $.
When calculating the decay amplitude of $\bar B_{s} \rightarrow V_1V_2$, where the intermediate particles consist of two vector mesons, the vector mediators exhibit three polarizations: longitudinal (L), normal (N), and transverse (T). The amplitude is also characterized by the polarization states of these vector mediators.

In this paper, we define the momenta of $\bar B_s$ meson and two vector mesons using light-cone coordinates. By analyzing the Lorentz structure, its decay amplitude is decomposed into the following form \cite{Chen:2002pz,Lu:2005be,Li:2004ti,Huang:2005if}:
\setlength\abovedisplayskip{0.4cm}
\setlength\belowdisplayskip{0.4cm}
\setlength{\jot}{8pt}
\begin{eqnarray}
	A^{(\sigma)}_1&=&\epsilon^*_{1\mu}\epsilon^*_{2\nu}(a g^{\mu \nu}+\frac{b}{M_{1} M_{2}} P^{\mu} P^{\nu}+\frac{i c}{M_{1} M_{2}} \epsilon^{ \mu \nu \alpha \beta} P_{1 \alpha} P_{2 \beta})\nonumber\\
   &=&M^2_{\bar B_{s}}A_{L}+M^2_{\bar B_{s}}A_{N}
	\epsilon^{*}_{2}(\sigma=T)\cdot\epsilon^{*}_{3}(\sigma=T) +i
	A_{T}\epsilon^{\alpha \beta\gamma \rho}
	\epsilon^{*}_{2\alpha}(\sigma)\epsilon^{*}_{3\beta}(\sigma)
	P_{2\gamma }P_{3\rho }\;,
\end{eqnarray}
The superscript $\sigma$ indicates the helicity states of two vector mesons, where L (T) respectively represent the longitudinal (transverse) components.  We can define the longitudinal $H_{0}$, transverse $H_{\pm}$ helicity amplitudes as :
\setlength\abovedisplayskip{0.3cm}
\setlength\belowdisplayskip{0.3cm}
\setlength{\jot}{8pt} 
\begin{eqnarray}
	&&H_{0}=M^{2}_{\bar B_{s}} {\cal A}_{L}\,,\nonumber\\
	&&H_{\pm}=M^{2}_{\bar B_{s}} {\cal A}_{N} \mp  M_{1} M_{2} \sqrt{r^{'2}-1}{\cal A}_{T}\,.
\end{eqnarray}
In the equation $r^{'2}=\frac{P_{V_1}\cdot P_{V_2}}{M_{1}M_{2}}=\frac{M^{2}_{\bar B_{s}}}{2M_{1}M_{2}}$, $P_{V_1}$ and $P_{V_2}$ denote the momenta of the two vector particles, while $M_{1}$ and $M_{2}$ represent their respective masses. These quantities satisfy the following equation:
\setlength\abovedisplayskip{0.5cm}
\setlength\belowdisplayskip{0.3cm}
\begin{equation}
	\sum_{\sigma}{ A_1}^{(\sigma)\dagger }{ A_1^{(\sigma)}}=|H_{0}|^{2}+|H_{+}|^{2} + | H_{-}|^{2}.
\end{equation}
 
There exists another equivalent set of definitions for the helicity amplitudes:
\setlength\abovedisplayskip{0.3cm}
\setlength\belowdisplayskip{0.3cm}
\begin{eqnarray}
	A_{0}&=&-\xi M^{2}_{\bar B_{s}}{\cal A}_{L}, \nonumber\\
	A_{\parallel}&=&\xi \sqrt{2}M^{2}_{\bar B_{s}}{\cal A}_{N}, \nonumber \\
	A_{\perp}&=&\xi M_{1} M_{2} \sqrt{2(r^{'2}-1)} {\cal A }_{T}\;,
	\label{ase}
\end{eqnarray}
 with $\xi$ the normalization factor to satisfy $	|A_{0}|^2+|A_{\parallel}|^2+|A_{\perp}|^2=1$, where the notations $A_{0}$, $A_{\parallel}$, $A_{\perp}$ denote the longitudinal, parallel, and perpendicular polarization amplitude.

In the process \( V_1 \rightarrow P_1P_2 \), the amplitude is given by \( A_{V_1 \rightarrow P_1P_2}^{\lambda} = g_{V_1} \epsilon(\lambda) \cdot (p_1 - p_2) \), and for the process \( V_2 \rightarrow P_3P_4 \), the amplitude is \( A_{V_2 \rightarrow P_3P_4}^{\gamma} = g_{V_2} \epsilon(\gamma) \cdot (q_1 - q_2) \). Here, \( g_V \) represents the effective coupling constant of the vector meson. By utilizing matrix elements that include the CKM matrix and incorporating Feynman diagrams, we can derive the amplitude form for the four-body decay:
\setlength\abovedisplayskip{0.3cm}
\setlength\belowdisplayskip{0.8cm}
\setlength{\jot}{11pt} 
\begin{eqnarray}
	\langle P_1P_2P_3P_4\left | H_{eff} \right | \bar{B}_{s}  \rangle
	&=&   \langle P_1P_2| V_1
	\rangle \langle P_3P_4  | V_2  \rangle
	\langle V_1V_2 \left | {\cal H}_{eff} \right | \bar{B}_{s} \rangle \nonumber\\
	&=&\frac{g_{V_1}\epsilon(\lambda)\cdot \left(p_1-p_2\right)
		g_{V_2}\epsilon(\gamma)\cdot \left(q_1-q_2\right) }{s_{V_1} s_{V_2}}\nonumber\\
	&&\epsilon^*_{1\mu}\epsilon^*_{2\nu}(a g^{\mu \nu}+\frac{b}{M_{1} M_{2}} P^{\mu} P^{\nu}+\frac{i c}{M_{1} M_{2}} \epsilon_{ \mu \nu \alpha \beta} P^{ \alpha}_1 P^{ \beta}_2).
\end{eqnarray}

\section{Computational Analysis of CP Violation under Resonance Effects}
	\subsection{The computational formalism for CP violation in the decay process ${\bar B_{s} \rightarrow [\phi(\rho^{0},\omega)\rightarrow K^{+} K^{-}] [{K^{*0}}\rightarrow K^{+}\pi^{-}]}$ }

\label{sec:spectra}
We first utilize the decay channel  ${\bar B_{s} \rightarrow [\phi(\rho^{0},\omega)\rightarrow K^{+} K^{-}] [{K^{*0}}\rightarrow K^{+}\pi^{-}]}$ as an illustrative example to investigate the CP violation, and present the corresponding decay diagram in Fig. 6.
\begin{figure}[h]
	\centering
	\includegraphics[height=10.67cm,width=16cm]{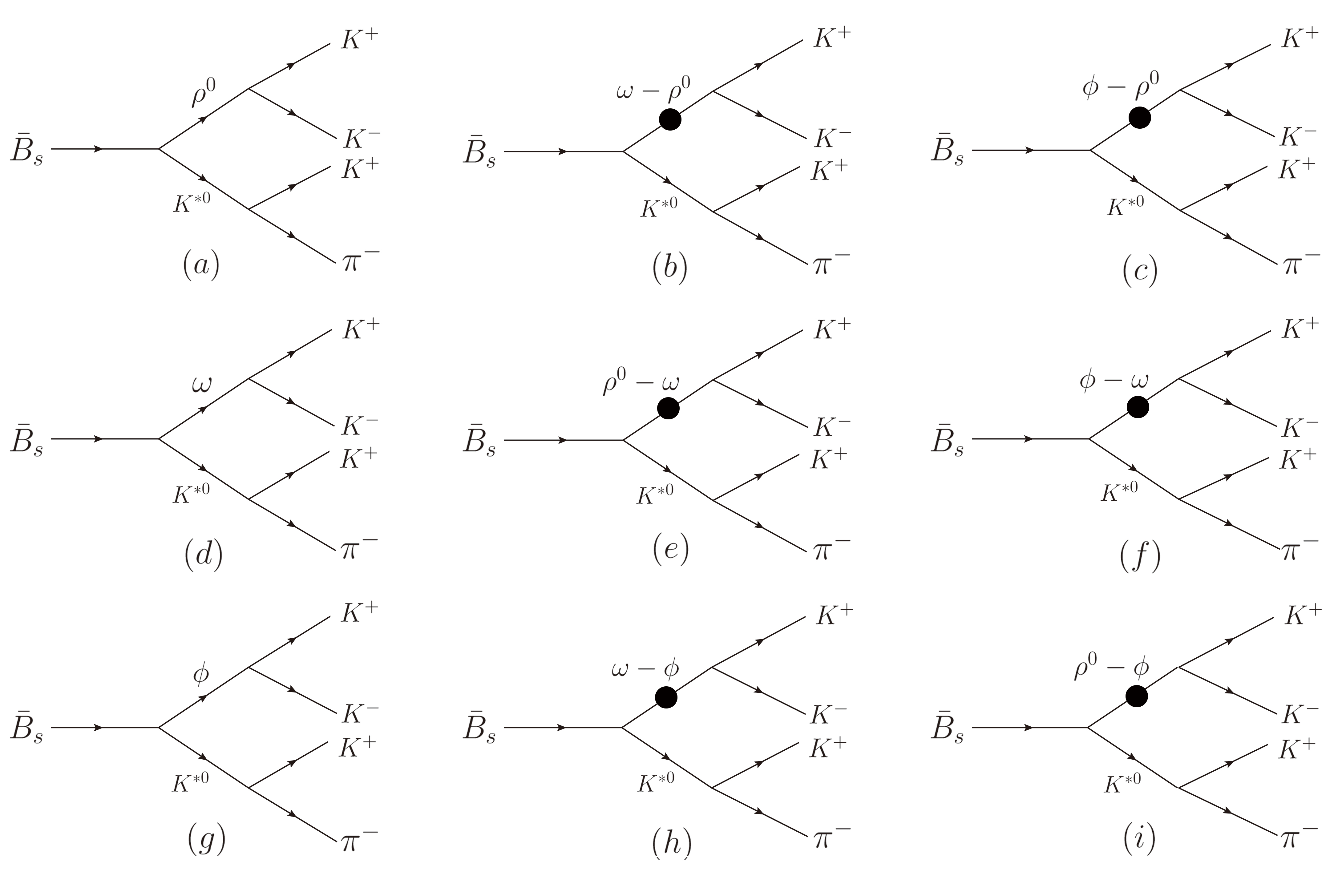}
	\caption{ The main contributing term in the amplitude of the ${\bar B_{s} \rightarrow [\phi(\rho^{0},\omega)\rightarrow K^{+} K^{-}] [{K^{*0}}\rightarrow K^{+}\pi^{-}]}$  process.}
	\label{fig5}
\end{figure}
In the aforementioned decay diagram, we have employed the quasi-two-body approach to effectively address the intricacies of the four-body decay process. In the decay process depicted in diagrams (a), (d) and (g) of Fig. 6,  the $K^{+} K^{-}$ pair is produced directly by $\rho^0$, $\omega$ and $\phi$ mesons,  while the $K^{+} \pi^{-}$ pair is produced directly by $K^{*0}$ meson. Furthermore, it is well-established that the production of KK mesons can also be attributed to various mixing effects. 
 In diagram (b), as compared to diagram (a), the production of the $K^{+} K^{-}$ pair can be attributed to the $\omega-\rho^0$ mixing effect, where the transition proceeds via $\omega \rightarrow \rho^0 \rightarrow K^{+} K^{-}$.
 The black dots represent the resonance effect in Fig. 6.  
The contribution of mixed resonance to the amplitude is relatively minor compared to that of diagrams (a), (d), and (g); However, it is still substantial enough to merit consideration. Diagrams (c), (e), (f), (h), and (i) display resonance effects analogous to those in diagram (b), specifically the $\omega-\rho^0$ ($\rho^0-\omega$), $\phi-\rho^0$ ($\rho^0-\phi$), and $\phi-\omega$ ($\omega-\phi$) resonances.

The amplitude of the four-body decay of  ${\bar B_{s} \rightarrow [\phi(\rho^{0},\omega)\rightarrow K^{+} K^{-}] [{K^{*0}}\rightarrow K^{+}\pi^{-}]}$ is expressed as follows:
\setlength\abovedisplayskip{0.3cm}
\setlength\belowdisplayskip{0.5cm}
\begin{equation}
	A=A_a+...+A_i,
\end{equation}
combined with the contribution of the decay process shown in Fig. 6. The total amplitude form of the decay process  ${\bar B_{s} \rightarrow [\phi(\rho^{0},\omega)\rightarrow K^{+} K^{-}] [{K^{*0}}\rightarrow K^{+}\pi^{-}]}$  in quasi-two-body form is obtained, as follows (In the following equation, for the sake of simplifying the expression, we have omitted the polarization vector and momentum terms.):
\setlength\abovedisplayskip{0.7cm}
\setlength\belowdisplayskip{0.7cm}
\begin{eqnarray}
	\begin{split}
		\left\langle K^{+} K^{-} K^{+} \pi^{-}\left|H_{eff}\right| \bar B_{s}\right\rangle=
		&\frac{g_{\rho^{0}\rightarrow K^+ K^-}g_{K^{*0}}}{s_{\rho^{0}}s_{K^{*0}}}A_{\rho^{0}K^{*0}}
		+\frac{g_{\rho^{0}\rightarrow K^+ K^-}g_{K^{*0}}}{s_{\omega}s_{\rho^{0}}s_{K^{*0}}}\widetilde{\Pi}_{\rho^{0}\omega}A_{\omega{K^{*0}}}
		+\frac{g_{\rho^{0}\rightarrow K^+ K^-}g_{K^{*0}}}{s_{\phi}s_{\rho^{0}}s_{K^{*0}}}\widetilde{\Pi}_{\rho^{0}\phi}A_{\phi{K^{*0}}}
		\\
		&
		+\frac{g_{\omega\rightarrow K^+ K^-}g_{K^{*0}}}{s_{\omega}s_{K^{*0}}}A_{\omega{K^{*0}}}
		+\frac{g_{\omega\rightarrow K^+ K^-}g_{K^{*0}}}{s_{\rho^{0}}s_{\omega}s_{K^{*0}}}\widetilde{\Pi}_{\omega\rho^{0}}A_{\rho^{0}{K^{*0}}}
		+\frac{g_{\omega\rightarrow K^+ K^-}g_{K^{*0}}}{s_{\phi}s_{\omega}s_{K^{*0}}}\widetilde{\Pi}_{\omega\phi}A_{\phi{K^{*0}}}
		\\
		&
		+\frac{g_{\phi\rightarrow K^+ K^-}g_{K^{*0}}}{s_{\phi}s_{K^{*0}}}A_{\phi{K^{*0}}}
		+\frac{g_{\phi\rightarrow K^+ K^-}g_{K^{*0}}}{s_{\omega}s_{\phi}s_{K^{*0}}}\widetilde{\Pi}_{\phi\omega}A_{\omega{K^{*0}}}
		+\frac{g_{\phi\rightarrow K^+ K^-}g_{K^{*0}}}{s_{\rho^{0}}s_{\phi}s_{K^{*0}}}\widetilde{\Pi}_{\phi\rho^{0}}A_{\rho^{0}{K^{*0}}}.
		\label{Htr}
	\end{split}	
\end{eqnarray}

The amplitudes of the decay processes $\bar B_s \rightarrow \rho^0 K^{*0}$, $\bar B_s \rightarrow \omega K^{*0}$, and $\bar B_s \rightarrow \phi K^{*0}$ are represented by $A_{\rho^{0}K^{*0}}$, $A_{\omega K^{*0}}$, and $A_{\phi K^{*0}}$, respectively. Here, $s_V$ denotes the inverse propagator of the vector meson $V (\rho^0, \omega, \phi, K^{*0})$ \cite{Chen:1999nxa,Wolfe:2009ts,Wolfe:2010gf}, while $g_V$ represents the coupling constant for the decay process $V \rightarrow P_1P_2$. Specifically, these coupling constants can be expressed as $\sqrt{2}g_{\rho^0 \rightarrow K^+ K^-} = \sqrt{2}g_{\omega \rightarrow K^+ K^-} = -g_{\phi \rightarrow K^+ K^-} = 4.54$ \cite{Bruch:2004py}, and $g_{K^{*0} \rightarrow K^+ \pi^-} = 4.59$ \cite{Cheng:2020ipp}.

According to Eq.(16), the amplitudes of the direct decays represented by diagrams (a), (d), and (g) in Fig. 6, denoted as $A_a$, $A_d$, and $A_g$, can be respectively expressed as follows:
\setlength\abovedisplayskip{0.7cm}
\setlength\belowdisplayskip{0.0cm}
\begin{eqnarray}
	A^{i=L,N,T}_a(\bar{B}_{s}^{0}\rightarrow\rho ^0 ( \rho ^0\rightarrow K^+K^-) K^{*0}(\ K^{*0}\rightarrow K^+\pi ^- ) )
	&=&\frac{G_Fg_{\rho }\epsilon \left( \lambda \right) \cdot \left( p_1-p_2 \right) g_{K^{*}}\epsilon \left( \gamma \right)\cdot  \left( q_1-q_2 \right)}{2 s_{\rho }s_{K^{*}}}A_{\rho^0K^{*0}}
	\nonumber\\
	&=&\frac{G_Fg_{\rho }\epsilon  \left( \lambda \right) \cdot \left( p_1-p_2 \right) g_{K^{*}}\epsilon \left( \gamma \right) \cdot \left( q_1-q_2 \right)}{2 s_{\rho }s_{K^{*}}}
	\nonumber\\
	&&
	\times \Bigg \{ V_{ub}V_{ud}^{*}\left[ f_{\rho}F_{B_s\rightarrow K^*}^{LL,i}\left( a_2 \right) +M_{B_s\rightarrow K^*}^{LL,i}\left( C_2 \right) \right]
	\nonumber\\
	&&+V_{tb}V_{td}^{*}[f_{\rho}F_{B_s\rightarrow K^*}^{LL,i}\left( -a_4+\frac{3}{2}a_7+\frac{3}{2}a_9+\frac{1}{2}a_{10} \right)
	\nonumber\\
	&&-M_{B_s\rightarrow K^*}^{LR,i}\left( -C_5+\frac{1}{2}C_7 \right) +M_{B_s\rightarrow K^*}^{LL,i}\left( -C_3+\frac{1}{2}C_9+\frac{3}{2}C_{10} \right)
	\nonumber\\
	&&-M_{B_s\rightarrow K^*}^{SP,i}\left( \frac{3}{2}C_8 \right) +f_{B_s}F_{ann}^{LL,i}\left( -a_4+\frac{1}{2}a_{10} \right)
	\nonumber\\
	&& -f_{B_s}F_{ann}^{SP,i}\left( -a_6+\frac{1}{2}a_8 \right) +M_{ann}^{LL,i}\left( -C_3+\frac{1}{2}C_9 \right)
	\nonumber\nonumber\\
	&&-M_{ann}^{LR,i}\left( -C_5+\frac{1}{2}C_7 \right)\bigg]\bigg\},   
\end{eqnarray}
\setlength\abovedisplayskip{0.0cm}
\setlength\belowdisplayskip{0.3cm}
\begin{eqnarray}
	A^{i}_d(\bar{B}_{s}^{0}\rightarrow \omega\left(\omega \rightarrow K^+K^- \right) K^{*0}(\ K^{*0}\rightarrow K^+\pi ^- ) )
		&=&\frac{G_Fg_{\omega}\epsilon \left( \lambda \right)\cdot  \left( p_1-p_2 \right) g_{K^{*}}\epsilon \left( \gamma \right) \cdot \left( p_1-p_2 \right)}{2s_{\omega}s_{K^{*}}}A_{\omega K^{*0}}
	\nonumber\\
	&=&\frac{G_Fg_{\omega}\epsilon \left( \lambda \right) \cdot \left( p_1-p_2 \right) g_{K^{*}}\epsilon \left( \gamma \right)\cdot  \left( p_1-p_2 \right)}{2s_{\omega}s_{K^{*}}}
	\nonumber\\
	&&
	\times\bigg\{V_{ub}V_{ud}^{*}\left[f_{\omega}F_{B_s\rightarrow K^*}^{LL,i}\left( a_2 \right) +M_{B_s\rightarrow K^*}^{LL,i}\left( C_2 \right) \right]
	\nonumber\\
	&&- V_{tb}V_{td}^{*}[ f_{\omega} F_{B_s\to K^*}^{LL,i}  \left(
	2a_{3}+a_{4}+2a_{5}
	+\frac{1}{2}a_{7}+\frac{1}{2}a_{9}-\frac{1}{2}a_{10}\right)
	\nonumber\\
	&&   + M_{B_s\to K^*}^{LL,i}\left(C_{3}+2C_{4}-\frac{1}{2}C_9
	+\frac{1}{2}C_{10}\right)
	\nonumber\\
	&&   - M_{B_s\to K^*}^{LR,i}\left(C_{5}-\frac{1}{2}C_{7}\right)
	- M_{B_s\to K^*}^{SP,i}\left(2C_{6}+\frac{1}{2}C_{8}\right)
	\nonumber\\
	&&
	+ f_{B_s}  F_{ann}^{LL,i}\left(a_{4} -\frac{1}{2}a_{10}\right)
	-f_{B_s} F_{ann}^{SP,i}\left(a_{6} -\frac{1}{2}a_{8}\right)
	\nonumber
	\\
	&&  + M_{ann}^{LL,i}\left(C_{3}-\frac{1}{2}C_{9}\right)
	- M_{ann}^{LR,i}\left(C_{5}-\frac{1}{2}C_{7}\right)\bigg]\bigg\},
\end{eqnarray}
\setlength\abovedisplayskip{0.0cm}
\setlength\belowdisplayskip{0.3cm} 
\begin{eqnarray}
	A^{i}_g(\bar{B}_{s}^{0}\rightarrow K^{*0}(\ K^{*0}\rightarrow K^+\pi ^- )  \phi\left( \phi \rightarrow K^+K^- \right) )
	&=& -\frac{G_Fg_{K^{*}}\epsilon(\lambda)\cdot \left(p_1-p_2\right)
		g_{\phi} \epsilon(\gamma)\cdot \left(q_1-q_2\right) }{\sqrt{2}s_{K^{*}} s_\phi}A_{\phi K^{*0}}
	\nonumber\\
	&=& -\frac{G_Fg_{K^{*}}\epsilon(\lambda)\cdot \left(p_1-p_2\right)
		g_{\phi} \epsilon(\gamma)\cdot \left(q_1-q_2\right) }{\sqrt{2}s_{K^{*}} s_\phi}
	\nonumber\\
	&&
	\times V_{tb}V_{td}^{*} \bigg[f_\phi F_{B_s\to K^*}^{LL,i}\left(
	a_{3}+a_{5}
	-\frac{1}{2}a_{7}-\frac{1}{2}a_{9}\right)
	\nonumber\\
	&&
	+   f_{K^*}F_{B_s\to \phi}^{LL,i}\left(
	a_{4} -\frac{1}{2}a_{10}\right)
	+  M_{B_s\to K^*}^{LL,i}\left(C_{4}-\frac{1}{2}C_{10}\right)
	\nonumber\\
	&&
	+  M_{B_s\to \phi}^{LL,i}\left(C_{3}-\frac{1}{2}C_9\right)
	- M_{B_s\to K^*}^{SP,i}\left(C_{6}-\frac{1}{2}C_{8}\right)
	\nonumber\\
	&&
	-M_{B_s\to \phi}^{LR,i}\left(C_{5}-\frac{1}{2}C_{7}\right)+  f_{B_s}
	F_{ann}^{LL,i}\left(a_{4} -\frac{1}{2}a_{10}\right)
	\nonumber	\\
	&&
	-f_{B_s}	F_{ann}^{SP,i}\left(a_{6} -\frac{1}{2}a_{8}\right)
	+  M_{ann}^{LL,i}\left(C_{3}-\frac{1}{2}C_{9}\right)
	\nonumber\\
	&&
	-M_{ann}^{LR,i}\left(C_{5}-\frac{1}{2}C_{7}\right)\bigg].
\end{eqnarray}

The coefficient $C_i$ represents the Wilson coefficient, while $a_i$ denotes the terms associated with Wilson coefficient. The variable $\epsilon$ indicates the polarization of the vector meson, $G_F$ is the Fermi constant, and $f_{\rho^{0}}$ refers to the decay constant of the $\rho^{0}$ meson \cite{Li:2006jv}. Additionally, $F_{\bar B_{s}\to V }^{LL}$ and $M_{ \bar B_{s}\to V}^{LL}$ represent factorizable emission diagrams and annihilation diagrams, respectively, whereas $F_{ann}^{LL}$ and $M_{ann}^{LL}$ denote non-factorizable emission diagrams and annihilation diagrams. The superscripts $LL$, $LR$, and $SP$ correspond to the contributions from the $(V-A)\otimes(V-A)$, $(V-A)\otimes(V+A)$, and $(S-P)\otimes(S+P)$ current operators, respectively \cite{Ali:2007ff}.
For the interference amplitudes caused by mixing represented in diagrams (b), (c), (e), (f), (h),  and (i) in Fig.6, taking diagrams (b) as an example, then $A_b$ can be expressed as:
\setlength\abovedisplayskip{0.3cm}
\setlength\belowdisplayskip{0.3cm}
\begin{eqnarray}
	A^{i}_b(\bar{B}_{s}^{0}\rightarrow \omega\rightarrow\rho^{0} \left( \rho^{0} \rightarrow K^+K^- \right) K^{*0}(\ K^{*0}\rightarrow K^+\pi ^- ) )
	&=&\frac{G_Fg_{\rho^0}\epsilon \left( \lambda \right)\cdot  \left( p_1-p_2 \right) g_{K^{*}}\epsilon \left( \gamma \right)\cdot  \left( p_1-p_2 \right)}{2 s_{\omega}s_{\rho^0}s_{K^{*}}}\widetilde{\Pi}_{\rho^{0}\omega}A_{\omega{K^{*0}}}
	\nonumber\\
	&=&\frac{G_Fg_{\rho^0}\epsilon \left( \lambda \right)\cdot  \left( p_1-p_2 \right) g_{K^{*}}\epsilon \left( \gamma \right)\cdot  \left( p_1-p_2 \right)}{2 s_{\omega}s_{\rho^0}s_{K^{*}}}\widetilde{\Pi}_{\rho^{0}\omega}
	\nonumber\\
	&&
	\times\bigg\{V_{ub}V_{ud}^{*}\left[f_{\omega}F_{B_s\rightarrow K^*}^{LL,i}\left( a_2 \right) +M_{B_s\rightarrow K^*}^{LL,i}\left( C_2 \right) \right]
	\nonumber\\
	&&- V_{tb}V_{td}^{*}[ f_{\omega} F_{B_s\to K^*}^{LL,i}  \left(
	2a_{3}+a_{4}+2a_{5}
	+\frac{1}{2}a_{7}+\frac{1}{2}a_{9}-\frac{1}{2}a_{10}\right)
	\nonumber\\
	&&   + M_{B_s\to K^*}^{LL,i}\left(C_{3}+2C_{4}-\frac{1}{2}C_9
	+\frac{1}{2}C_{10}\right)
	\nonumber\\
	&&   - M_{B_s\to K^*}^{LR,i}\left(C_{5}-\frac{1}{2}C_{7}\right)
	- M_{B_s\to K^*}^{SP,i}\left(2C_{6}+\frac{1}{2}C_{8}\right)
	\nonumber\\
	&&
	+ f_{B_s}  F_{ann}^{LL,i}\left(a_{4} -\frac{1}{2}a_{10}\right)
	-f_{B_s} F_{ann}^{SP,i}\left(a_{6} -\frac{1}{2}a_{8}\right)
	\nonumber
	\\
	&&  + M_{ann}^{LL,i}\left(C_{3}-\frac{1}{2}C_{9}\right)
	- M_{ann}^{LR,i}\left(C_{5}-\frac{1}{2}C_{7}\right)\bigg]\bigg\}.
\end{eqnarray}

For the four-body decay, the polarization fractions $f_i(i=0,\parallel,\perp)$ are defined as follows:
\setlength\abovedisplayskip{0.3cm}
\setlength\belowdisplayskip{0.3cm}
\begin{eqnarray}
	f_i=\frac{|A_i|^2}{|A_0|^2+|A_\parallel|^2+|A_\perp|^2}\;.
\end{eqnarray}
$f_0+f_{\parallel}+f_{\perp}=1$ that satisfies the normalization condition.

The differential parameters of CP violation are calculated in the following form:
   \setlength\abovedisplayskip{0.5cm}
\setlength\belowdisplayskip{0.5cm}
\begin{equation}
	\label{cp31}
	A_{CP}=\frac{\left| A \right|^2-\left| \overline{A} \right|^2}{\left| A \right|^2+\left| \overline{A} \right|^2}.
\end{equation}

In recent years, research data on CP violation in multibody decays of B mesons have been accumulated through collaborative experiments such as LHCb \cite{LHCb:2018oeb,LHCb:2019maw,LHCb:2023sim}, Belle \cite{Belle:2009roe,Belle:2010uya}, and BaBar \cite{BaBar:2007wwj,BaBar:2008zea}. These studies primarily focus on the invariant mass region of quasi-two-body decays of B mesons. To facilitate comparison with experimental results, we compute the $A_{C P}^{\Omega}$-integral, which accounts for both resonant and non-resonant contributions within a specific region $\Omega$. By integrating the numerator and denominator of $A_{C P}$ over region $\Omega$, we obtain the regional integral CP violation. Its form is as follows:
  \setlength\abovedisplayskip{0.9cm}
\setlength\belowdisplayskip{1cm}
\begin{equation}
	A_{C P}^{\Omega}=\frac{\int \mathrm{d} \Omega\left(|A|^{2}-|\overline{A}|^{2}\right)}{\int  \mathrm{d} \Omega\left(|A|^{2}+|\overline{A}|^{2}\right)}.
\end{equation}

\subsection{The computational formalism for CP violation in the decay process $\bar B_{s} \rightarrow [\rho^{0} (\omega, \phi) \rightarrow \pi^{+} \pi^{-}] [K^{*0} \rightarrow K^{+} \pi^{-}]$
}

We further investigated the CP violation in the decay process $\bar B_{s} \rightarrow [\rho^{0} (\omega,\phi) \rightarrow \pi^{+} \pi^{-}] [K^{*0} \rightarrow K^{+} \pi^{-}]$, and the corresponding decay diagrams are illustrated in Fig. 7 and Fig. 8.
Since the decay modes of $\rho^{0}\rightarrow \pi^+\pi^-$ and $K^{*0}\rightarrow K^+\pi^-$ have branching ratios of $100\%$, the dominant contribution to the amplitude in the $\bar B_{s} \rightarrow [\rho^{0} (\omega,\phi)\rightarrow\pi^{+} \pi^{-}] [K^{*0}\rightarrow K^{+}\pi^{-}]$ process originates from diagram (a) in Fig. 7. Diagrams (b) and (c) of Fig. 7 also contribute significantly, representing the mixed resonance contributions to the amplitude through the processes $\omega-\rho^0$ $(\omega\rightarrow\rho^0\rightarrow \pi^{+} \pi^{-})$ and $\phi-\rho^0$ $(\phi\rightarrow\rho^0\rightarrow \pi^{+} \pi^{-})$. These resonance processes result in mixing amplitudes.
\begin{figure}[h]
	\centering
	\includegraphics[height=3.56cm,width=16cm]{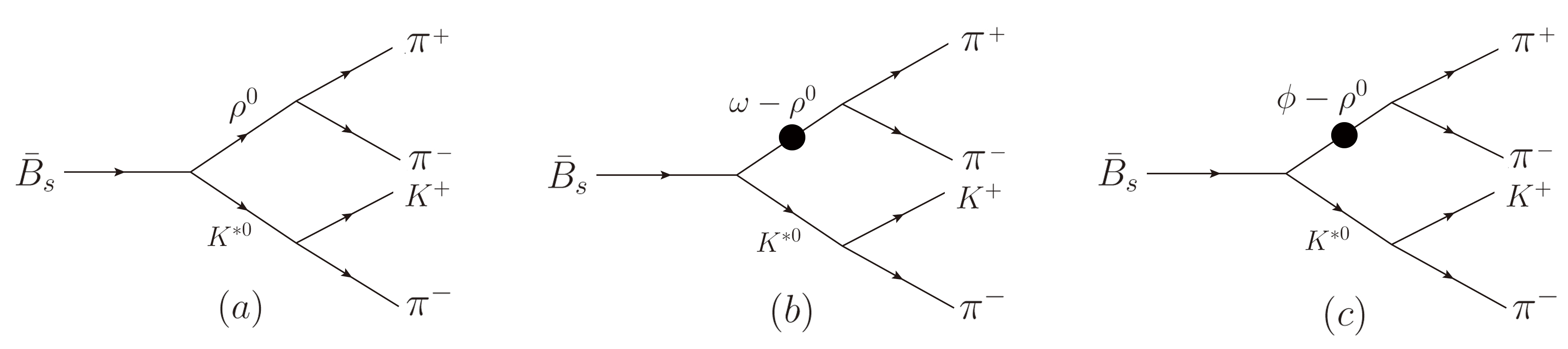}
	\caption{   The main contributing term in the amplitude of the $\bar B_{s} \rightarrow[\rho^{0} ( \omega,\phi)\rightarrow\pi^{+}\pi^{-}] [{K^{*0}}\rightarrow K^{+}\pi^{-}] $ decay process.}
	\label{fig6}
\end{figure}

However, in diagrams (d) and (g) of Fig. 8, the CP violation arising from isospin symmetry breaking in the $\omega\rightarrow\pi^{+} \pi^{-}$ and $\phi\rightarrow\pi^{+} \pi^{-}$ decay processes are higher-order terms ($\varepsilon_1$) and can thus be neglected. Additionally, the mixed parameters $\widetilde{\Pi}_{\rho^0\omega}$, $\widetilde{\Pi}_{\rho^0\phi}$, and $\widetilde{\Pi}_{\omega\phi}$ are small quantities of first-order approximation ($\varepsilon_2$). In diagrams (e), (f), (h), and (i), the product $\varepsilon_1 \varepsilon_1$ represents a higher-order small quantity. Therefore, the contributions from diagrams (d) through (i) of Fig. 8 can be disregarded. 
\begin{figure}[h]
	\centering
	\includegraphics[height=6.69cm,width=16cm]{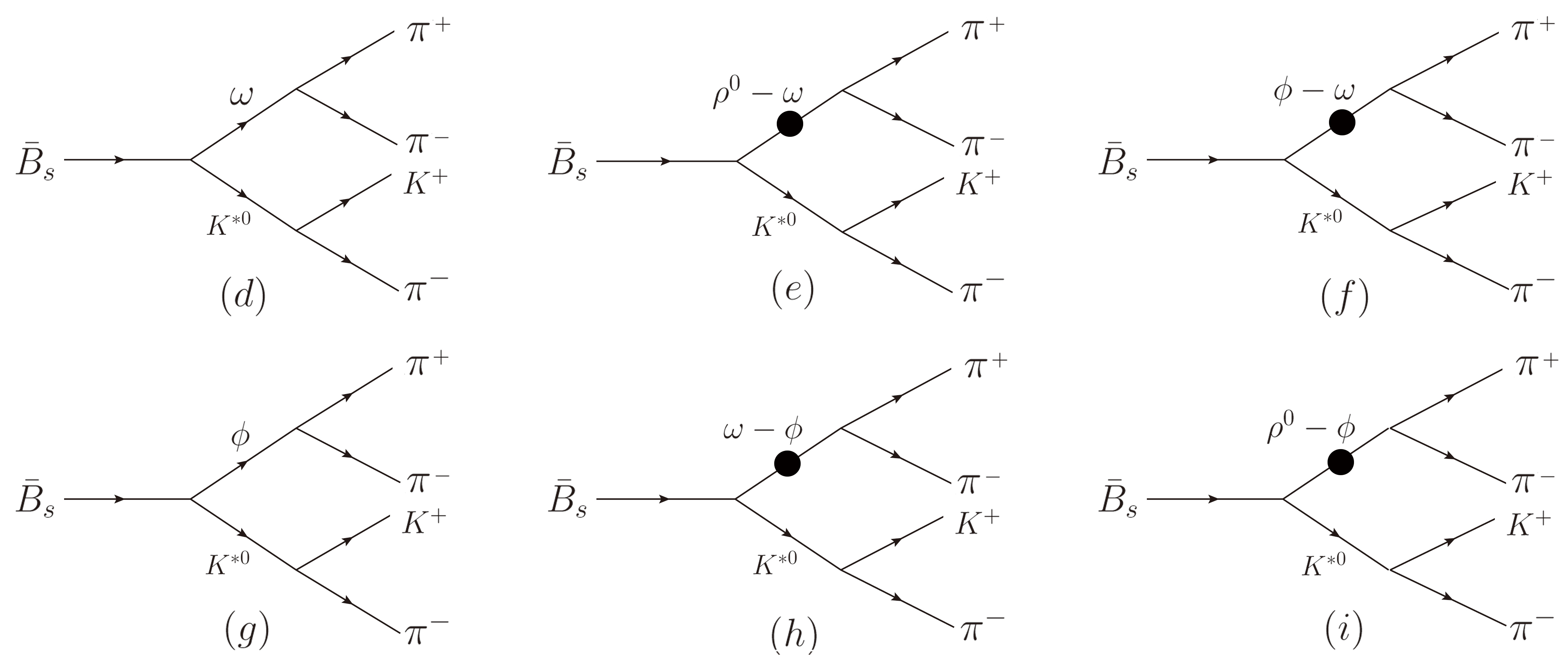}
	\caption{ The contribution terms of amplitudes that can be neglected in the $\bar B_{s} \rightarrow [\rho^{0} ( \omega,\phi)\rightarrow\pi^{+} \pi^{-}] [{K^{*0}}\rightarrow K^{+}\pi^{-}] $ decay process.  }
	\label{fig6}
\end{figure}

Consequently, the $\bar B_{s} \rightarrow [\rho^{0} (\omega,\phi)\rightarrow\pi^{+} \pi^{-}] [{K^{*0}}\rightarrow K^{+}\pi^{-}]$ decay process primarily involves contributions from diagrams (a), (b), and (c) of Fig. 7, as indicated by the aforementioned considerations. The total amplitude associated with diagrams (a), (b), and (c) of Fig. 7 can be expressed as:
  \setlength\abovedisplayskip{0.1cm}
\setlength\belowdisplayskip{0.1cm}
\begin{equation}
	A=A_a+A_b+A_c.
\end{equation}
Combined with the contribution of the decay process in the above consideration, the total amplitude form of the decay process of $\bar B_{s} \rightarrow [\rho^{0} ( \omega,\phi)\rightarrow\pi^{+} \pi^{-}] [{K^{*0}}\rightarrow K^{+}\pi^{-}] $ is obtained (Similarly, in the following equation, we have omitted the polarization vector and momentum terms.):
 \setlength\abovedisplayskip{0.4cm}
\setlength\belowdisplayskip{0.0cm}
\begin{eqnarray}
		\left\langle \pi^{+} \pi^{-} K^{+} \pi^{-}\left|H_{eff}\right| \bar B_{s}\right\rangle=
		\frac{g_{\rho^{0} \rightarrow\pi^{+}\pi^{-}}g_{K^{*0}}}{s_{\rho^{0}}s_{K^{*0}}}A_{\rho^{0}K^{*0}}
		+\frac{g_{\rho^{0} \rightarrow\pi^{+}\pi^{-}}g_{K^{*0}}}{s_{\omega}s_{\rho^{0}}s_{K^{*0}}}\widetilde{\Pi}_{\rho^{0}\omega}A_{\omega{K^{*0}}}
		+\frac{g_{\rho^{0} \rightarrow\pi^{+}\pi^{-}}g_{K^{*0}}}{s_{\phi}s_{\rho^{0}}s_{K^{*0}}}\widetilde{\Pi}_{\rho^{0}\phi}A_{\phi{K^{*0}}}.\nonumber\\
\end{eqnarray}
Here the coupling constant of the $\rho^{0} \rightarrow\pi^{+}\pi^{-}$process is $g_{\rho^{0} \rightarrow\pi^{+}\pi^{-}}=6.00$ \cite{Cheng:2020ipp}. 
The amplitudes $A_a$, $A_b$, and $A_c$ follow the similar formalism as presented in Eqs. (19)-(22). The amplitudes ($A_{\rho^{0}K^{*0}}$, $A_{\omega K^{*0}}$, and $A_{\phi K^{*0}}$) of the two-body processes $\bar B_s \rightarrow \rho^0 K^{*0}$, $\bar B_s \rightarrow \omega K^{*0}$, and $\bar B_s \rightarrow \phi K^{*0}$ are the same, and the corresponding parameters need to be modified.
The CP violation differential parameter $A_{CP}$ , the polarization fractions $f_i(i=0,\parallel,\perp)$ and the integrated CP violation over the region $\Omega$, denoted as $A_{CP}^{\Omega}$, for the decay process $\bar B_{s} \rightarrow \rho^{0} (\omega,\phi) K^{*0} \rightarrow \pi^{+} \pi^{-} K^{+} \pi^{-}$ are consistent with Eqs. (23), (24) and (25).

\section{INPUT PARAMETER}

The elements of the CKM matrix can be parameterized using the Wolfenstein parameters $\lambda$, $A$, $\bar{\rho}$, and $\bar{\eta}$ \cite{Cabibbo:1963yz,Kobayashi:1973fv,Wolfenstein:1983yz}. Specifically, $V_{ub} V_{ud}^* = A \lambda^3 (\bar{\rho} - i \bar{\eta}) (1 - \frac{\lambda^2}{2})$ and $V_{tb} V_{td}^* = A \lambda^3 (1 - \bar{\rho} + i \bar{\eta})$. The most recent values for these parameters are provided in Ref. \cite{ParticleDataGroup:2024cfk}:
\setlength\abovedisplayskip{0.1cm}
\setlength\belowdisplayskip{0.2cm}
\setlength{\jot}{3pt} 
\begin{eqnarray}
	\begin{split}
		&& \lambda=0.22650\pm0.00048,\quad A=0.790^{+0.017}_{-0.012}, \\
		&& \bar{\rho}=0.141_{-0.017}^{+0.016},\hspace{0.8cm}\quad
		\bar{\eta}=0.357\pm0.011,
	\end{split}
\end{eqnarray}
where $\bar{\rho}=\rho(1-\frac{\lambda^2}{2}),\bar{\eta}=\eta(1-\frac{\lambda^2}{2})$.
 
The central values of the parameters used in the calculation are given in the Table I \cite{Li:2006jv,ParticleDataGroup:2024cfk}.
\begin{table}[h]
	\caption{Parameters used in calculation}
	\centering
	\begin{tabular*}{16.5cm}{@{\extracolsep{\fill}}l||cccc}
		
		\hline\hline\\
		\text{Parameters}&$$&\text{Input data}\\[1ex]\hline\hline\\
		\text{Mass ($\text{GeV}$)}
		&$m_{B_s}=5.36677$  &$m_{W}=80.385$  &$m_{\rho^0}=0.7753$ &$m_{\omega}=0.7827$   \\[1ex]
			\text{}
		               &$m_{\phi}=1.0195$ &$m_{K^{*0}}=0.8917$ &$m_{\pi^{0}}=0.135$ &$m_{K^{0}}=0.4976$ \\[1ex]  
		 \text{}
		   &$m_{\pi^{\pm}}=0.1396$   &$m_{K^{\pm}}=0.4937$\\[1ex]\\
		\text{Decay constants ($\text{GeV}$) }
		& $f_{\rho^{0}}=0.209$ & $f_{\omega}=0.195$  &$f_{\phi}=0.231$ &$f_{K^{*0}}=0.217$\\[1ex]
		\text{ }
		&$f_{\rho^{0}}^T=0.169$&$f_{\omega}^T=0.145$&$f_{\phi}^T=0.200$&$f_{K^{*0}}^T=0.185$\\[1ex]  
		\text{}
		  &$f_{\bar{B_s}}$= $0.23$\\[1ex] \\
			\text{Decay width ($\text{GeV}$)}
			& $\Gamma_{\rho} = 0.15 $ & $\Gamma_{\omega} = 8.68 \times 10^{-3}$ &$ \Gamma_{\phi} = 4.25 \times 10^{-3} $   & $ \Gamma_{{K^{*0}}} = 4.73 \times 10^{-2}$ \\[1ex]\\
			\text{Lifetime ($\text{ps}$)}
			& $\tau_{B_s} = 1.51 $ & $ $   & $ $ \\[1ex]
				\hline\hline
			\end{tabular*}
		\end{table}

		\section{NUMERICAL RESULT ANALYSIS}

		Based on the discussion in the previous chapter, we  have obtained  the relationship between the $A_{CP}$ and the invariant mass $\sqrt{s}$ in the decay process.
		In the calculation, we find that the contribution of  $g_{K^{*0}}$ and $s_{K^{*0}}$ in the $\sqrt{s_2}$ (the invariant mass of $K^{+}\pi^{-}$) direction is completely eliminated, so we mainly analyze the relationship between $A_{CP}$ and  $\sqrt{s_1}$ (the invariant mass of $K^{+}K^{-}$ or $\pi^{+}\pi^{-}$). 	
		\begin{table}[!htbh]
			\caption{ Direct $A_{CP}$ values (in units of $\%$) of $\bar B_{s} \rightarrow {(K^{+} K^{-})( K^{+}\pi^{-})} $ and $\bar B_{s} \rightarrow {(\pi^{+} \pi^{-})( K^{+}\pi^{-})} $ decay processes in different intermediate states.}
			\begin{ruledtabular}  
				\setlength{\extrarowheight}{12.5pt}
					\setlength{\tabcolsep}{2mm}{ \begin{tabular}[t]{lcccc }
							Decay Modes      & ${\cal A}^{\rm CP}_{0}$       &${\cal A}^{\rm CP}_{\|}$  & ${\cal A}^{\rm CP}_{\bot}$ & ${\cal A}^{\rm CP}$
							\\\hline
							$\bar{B}_{s}^{0}\rightarrow \rho^0\left(\rho^0 \rightarrow K^+K^- \right)K^{*0} (\ K^{*0}\rightarrow K^+\pi ^-)$         &$63.57^{\pm1.46}_{\pm8.45}$ &$71.01^{\pm0.73}_{\pm6.10}$ &$25.15^{\pm0.14}_{\pm5.34}$ &$53.24^{\pm0.57}_{\pm15.73}$\\
							$\bar{B}_{s}^{0}\rightarrow \omega\left(\omega \rightarrow K^+K^- \right)K^{*0} (\ K^{*0}\rightarrow K^+\pi ^-)$  &$-77.38^{\pm 1.97}_{\pm7.92}$   &$53.44^{\pm0.92}_{\pm5.23}$    &$78.43^{\pm1.34}_{\pm8.11}$     &$-60.31^{\pm1.82}_{\pm13.12}$  \\
							$\bar{B}_{s}^{0}\rightarrow \phi\left(\phi \rightarrow K^+K^- \right)K^{*0} (\ K^{*0}\rightarrow K^+\pi ^-)$   &$\cdots$&$\cdots$&$\cdots$&$0$\\ \hline
							$\bar{B}_{s}^{0}\rightarrow\rho^0\left(\rho^0 \rightarrow \pi^+\pi^- \right) K^{*0}(\ K^{*0}\rightarrow K^+\pi ^-)$         &$63.58^{\pm1.44}_{\pm8.45}$ &$71.01^{\pm0.72}_{\pm6.11}$ &$25.14^{\pm0.14}_{\pm5.34}$ &$53.24^{\pm0.57}_{\pm15.73}$\\
							$\bar{B}_{s}^{0}\rightarrow\omega\left(\omega \rightarrow \pi^+\pi^- \right) K^{*0}(\ K^{*0}\rightarrow K^+\pi ^-)$  &$-77.39^{\pm1.98}_{\pm7.92}$   &$53.44^{\pm0.92}_{\pm5.23}$    &$78.44^{\pm1.35}_{\pm8.11}$     &$-60.31^{\pm1.82 }_{\pm13.12}$   \\
							$\bar{B}_{s}^{0}\rightarrow\phi\left(\phi \rightarrow \pi^+\pi^- \right) K^{*0}(\ K^{*0}\rightarrow K^+\pi ^-)$   &$\cdots$&$\cdots$&$\cdots$&$0$\\
					\end{tabular}}
			
			\end{ruledtabular}
		\end{table}
				
				
	
		In the Table II, we give the results of the four-body decay without the mixing effect of the intermediate state, which provides a reference for the experiment. 		
		The results indicate that the direct CP violation ($A_{CP}$) values for the decay processes $\bar{B}_{s}^{0}\rightarrow \phi\left(\phi \rightarrow K^+K^- \right) K^{*0}(\ K^{*0}\rightarrow K^+\pi ^-)$ and $\bar{B}_{s}^{0}\rightarrow \phi\left(\phi \rightarrow \pi^+\pi^- \right) K^{*0}(\ K^{*0}\rightarrow K^+\pi ^-)$ are both $0\%$. This is consistent with theoretical expectations, as these decays are dominated by penguin diagrams only.		
		The central value of $A_{CP}$  for $\bar{B}_{s}^{0}\rightarrow \rho^0\left(\rho^0 \rightarrow K^+K^- \right) K^{*0}(\ K^{*0}\rightarrow K^+\pi ^-)$ and $\bar{B}_{s}^{0}\rightarrow \omega\left(\omega \rightarrow K^+K^- \right)K^{*0} (\ K^{*0}\rightarrow K^+\pi ^-)$ are $53.24\% $ and $-60.31\%$, respectively. For the processes of $\bar{B}_{s}^{0}\rightarrow \rho^0\left(\rho^0 \rightarrow K^+K^- \right)K^{*0} (\ K^{*0}\rightarrow K^+\pi ^-)$ and $\bar{B}_{s}^{0}\rightarrow\rho^0\left(\rho^0 \rightarrow \pi^+\pi^- \right) K^{*0}(\ K^{*0}\rightarrow K^+\pi ^-)$, the CP-violation values of the four-body systems are almost the same. This is because in the four-body kinematics processes, the weak phases of the two processes are the same, and the strong phases cancel each other out, with the momentum having a negligible impact on the results. The same holds true for the $\bar{B}_{s}^{0}\rightarrow \omega\left(\omega \rightarrow K^+K^- \right)K^{*0} (\ K^{*0}\rightarrow K^+\pi ^-)$ and $\bar{B}_{s}^{0}\rightarrow\omega \left(\omega \rightarrow \pi^+\pi^- \right)K^{*0} (\ K^{*0}\rightarrow K^+\pi ^-)$ processes.
		  
	After the introduction of the mixing mechanism, the phenomenon of CP violation has undergone significant changes. When the invariant mass of the $K^{+} K^{-}$ pair of the $\bar B_{s} \rightarrow {(K^{+} K^{-}) (K^{+}\pi^{-})} $ decay channel is near the resonance range of $\phi$  ,  the CP violation ranges from $74.98\pm0.44\pm18.86\% $ to $ -81.73\pm0.35\pm12.56\%$.		
		For the decay process of $\bar B_{s} \rightarrow {\pi^{+} \pi^{-} K^{+}\pi^{-}}$, when the $\rho^{0}$ and $\omega$ resonances fall within the resonance region, the peak value of CP violation can reach $-71.04\pm0.77\pm11.98\%$.	 Based on the above results, we can find that the addition of the vector particle mixing mechanism significantly changes CP violation.

		 In order to better understand the regional CP violation and provide theoretical predictions for future experiments, we provide the ${A}^{\Omega} _{\mathrm{CP}}$ integral in the decay process. The numerical results are given in Table III and Table IV.
	\setlength{\extrarowheight}{1.8pt}
\begin{table}[h]	
	{\renewcommand
		\scalebox{1}
		\renewcommand{\arraystretch}{2} 
		\setlength{\tabcolsep}{10mm}
		\caption{The peak regional integral of ${A}^{\Omega} _{\mathrm{CP}}$ from different resonance rangs for $\bar B_{s} \rightarrow {(K^{+} K^{-}) (K^{+}\pi^{-})} $ decay processes.}}
	\begin{tabular*}{13.5cm}{@{\extracolsep{\fill}}lcccc}
		\hline\hline\\
		\text{Decay channel}&\text{Different resonance effect}&\text{$\sqrt{s_1}=$ $\mathrm{0.99-1.20}(GeV)$}\\[1ex]\hline\hline
		\\\text{$\bar B_{s} \rightarrow  (K^{+}K^{-}) (K^{+}\pi^{-}) $}&$\phi-\rho-\omega$&$\mathrm{-0.291\pm0.078\pm0.198}$\\[1ex]
		\text{}&$\phi-\rho$&$\mathrm{-0.006\pm0.001\pm0.012}$\\[1ex]
		\text{}&$\phi-\omega$&$\mathrm{-0.021\pm0.009\pm0.034}$\\[1ex]
		\text{}&$\rho-\omega$&$\mathrm{0.016\pm0.029\pm0.154}$\\[1ex] \hline\hline
	\end{tabular*}
\end{table}

In Table III, we provide the regional integrals ${A}^{\Omega} _{\mathrm{CP}}$ of peak values for different resonance ranges during the decay of $\bar B_{s} \rightarrow {(K^{+} K^{-}) (K^{+}\pi^{-})} $.
Taking into account the threshold effects of the KK pair and in order to comprehensively analyze the trend of CP violation, we have expanded the integration range from $0.99$ $GeV$ to $1.20$ $GeV$.
 In this decay channel, the resonance effect of three-particle mixing has more significant effect on ${A}^{\Omega} _{\mathrm{CP}}$ value than that of two-particle mixing. In the case of  $\rho^0-\omega$  situation, the result symbol of  ${A}^{\Omega} _{\mathrm{CP}}$ obtained is opposite to that when $\phi$ mesons are included. The penguin diagram contribution of process $\bar{B}_{s}^{0}\rightarrow \phi\left(\phi \rightarrow \pi^+\pi^- \right) K^{*0}(\ K^{*0}\rightarrow K^+\pi ^-)$ plays  major role in the whole.
	\setlength{\extrarowheight}{1.8pt}
\begin{table}[h]
	{\renewcommand
		\scalebox{1.5}
		\renewcommand{\arraystretch}{3} 
		\setlength{\tabcolsep}{8mm}
		\caption{The peak regional integral of ${A}^{\Omega} _{\mathrm{CP}}$ from different resonance rangs for $\bar B_{s} \rightarrow {(\pi^{+} \pi^{-}) (K^{+}\pi^{-})} $ decay processes}
		\centering}	
	\begin{tabular*}{13.5cm}{@{\extracolsep{\fill}}lcccccc}
		\hline\hline\\
		\text{Decay channel}&\text{Different resonance effect}&\text{$\sqrt{s_{1}}=$ $\mathrm{0.70-1.10}(GeV)$}&\\[1ex] \hline\hline
		\\\text{$\bar B_{s} \rightarrow {(\pi^{+} \pi^{-}) (K^{+}\pi^{-})} $}&$\rho-\omega-\phi$&$\mathrm{-0.537\pm0.020\pm0.135}$\\[1ex]
		\text{}&$\rho-\phi$&$\mathrm{-0.532\pm0.006\pm0.085}$\\[1ex]
		\text{}&$\rho-\omega$&$\mathrm{-0.537\pm0.021\pm0.134}$\\[1ex]
		\text{}&$\omega-\phi$&$\mathrm{-0.302\pm0.012\pm0.192}$\\[1ex] \hline\hline
		
	\end{tabular*}
\end{table}

For decay process $\bar B_{s} \rightarrow {(\pi^{+} \pi^{-}) (K^{+}\pi^{-})} $ in Table IV, we also provide regional integrals ${A}^{\Omega} _{\mathrm{CP}}$ of peaks in different resonance ranges in the integration interval $0.70$ $GeV$ to $1.10$ $GeV$. 
By comparing the results of different intermediate states in Table II, which include $\rho^0-\omega-\phi$ mixing, $\rho^0-\phi$ mixing, $\rho^0-\omega$ mixing, and $\omega-\phi$ mixing, it is evident that the mixed resonance processes alter CP violation both in terms of signs and magnitudes.
This decay process, $\bar{B}_{s}^{0}\rightarrow\phi \left(\phi \rightarrow \pi^+\pi^- \right) K^{*0}(\ K^{*0}\rightarrow K^+\pi ^-)$, is predominantly governed by penguin diagram contributions, with negligible tree diagram contributions. Consequently, CP violation is more pronounced in decay processes that involve intermediate states with $\rho^{0}$ particles. In comparison to $\omega-\phi$ mixing, the CP violation arising from $\rho^0-\phi$ mixing is significantly more evident.

\begin{figure}
	\centering
	\begin{minipage}[h]{0.51\textwidth}
		\centering
		\includegraphics[height=5cm,width=9.59cm]{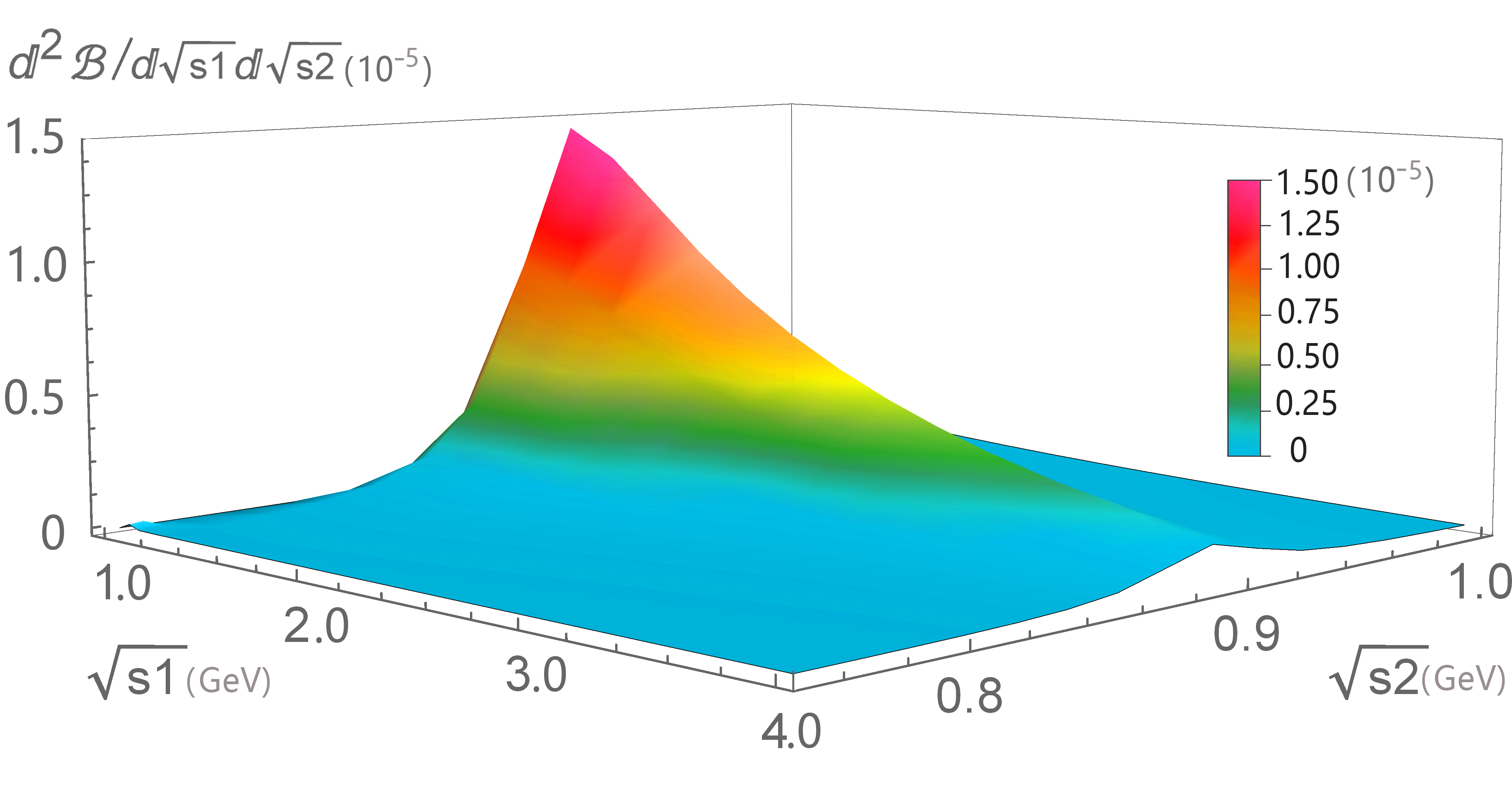}
		\caption{The differential branching ratio plot for the quasi-two-body decay process  $\bar B_{s} \rightarrow [\phi(\rho^{0},\omega)\rightarrow K^{+}K^{-}][K^{*0}\rightarrow K^{+}\pi^{-}]$ .}
		\label{fig2}
	\end{minipage}
	\quad
	\begin{minipage}[h]{0.45\textwidth}
		\centering
		\includegraphics[height=5cm,width=5.155cm]{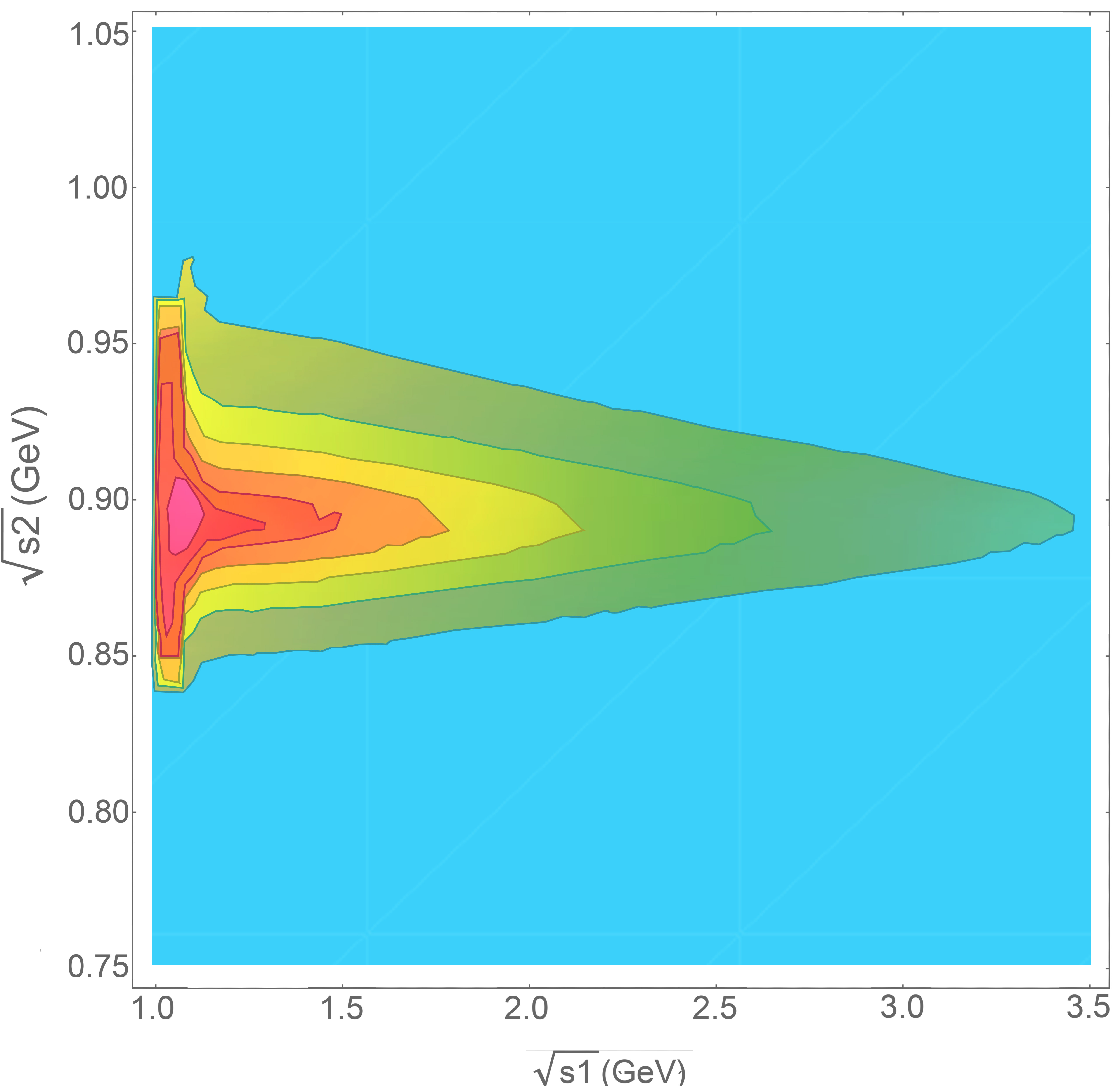}
		\caption{The plane mapping plot of the differential branching ratios for the decay process $\bar B_{s} \rightarrow [\phi(\rho^{0},\omega)\rightarrow K^{+}K^{-}][K^{*0}\rightarrow K^{+}\pi^{-}]$.}		
		\label{fig3}
	\end{minipage}
\end{figure}

\begin{figure}
	\centering
	\begin{minipage}[h]{0.51\textwidth}
		\centering
		\includegraphics[height=5cm,width=9.6cm]{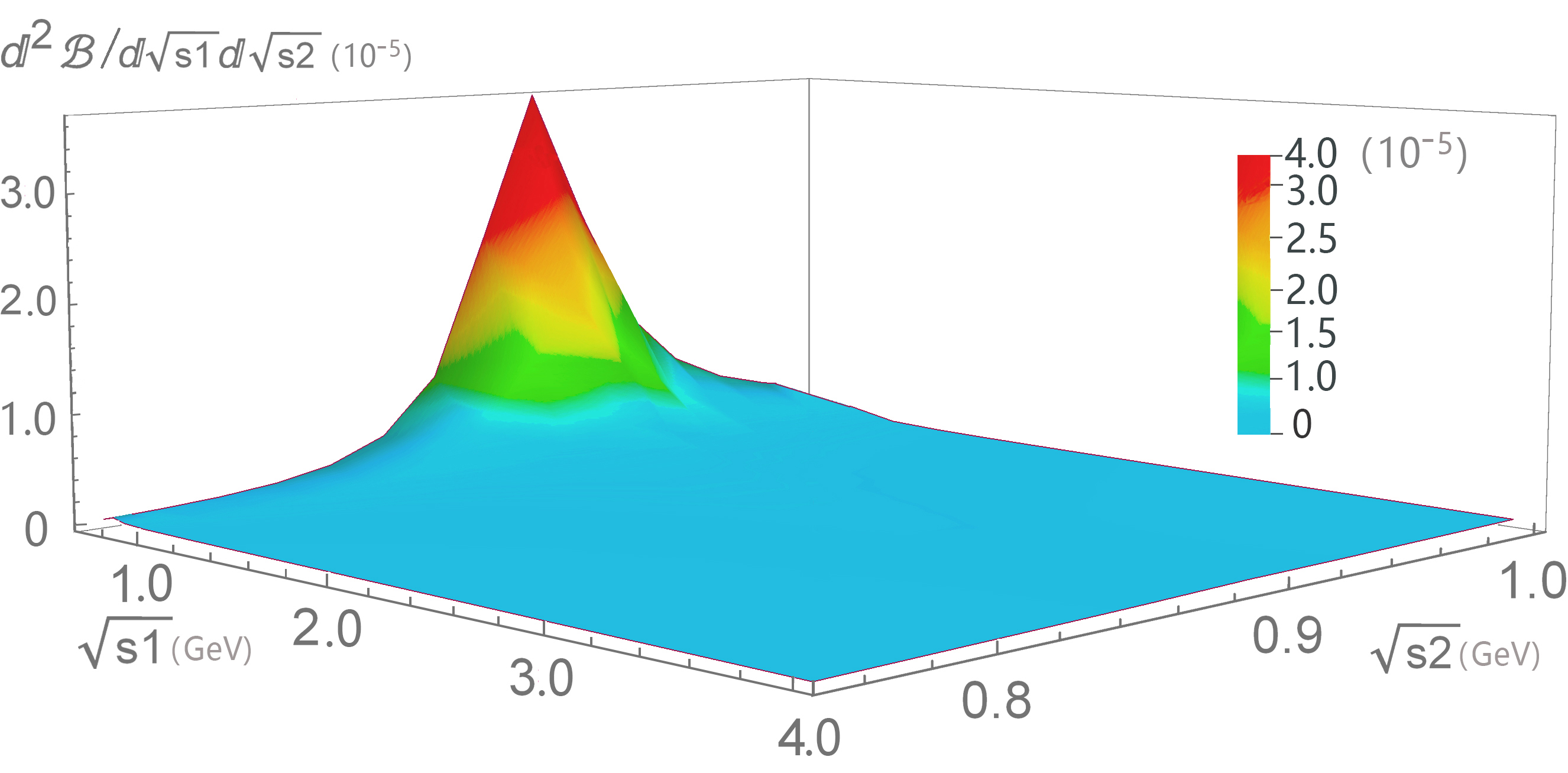}
		\caption{The differential branching ratio plot for the quasi-two-body decay process  $\bar B_{s} \rightarrow [\rho^{0}(\omega,\phi)\rightarrow (\pi^{+}\pi^{-})][K^{*0}\rightarrow K^{+}\pi^{-}]$ .}
		\label{fig2}
	\end{minipage}
	\quad
	\begin{minipage}[h]{0.45\textwidth}
		\centering
		\includegraphics[height=5cm,width=5.116625cm]{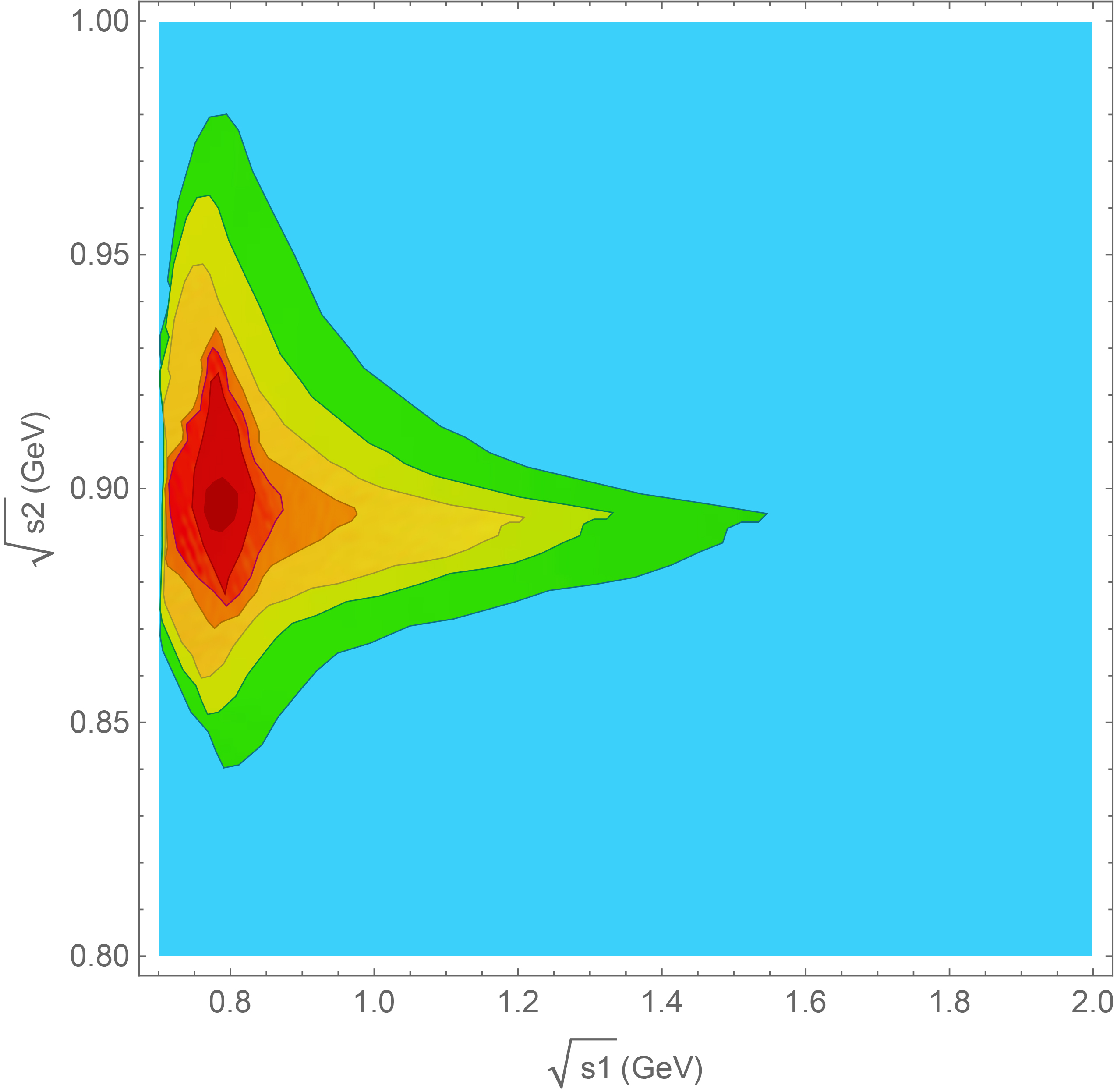}
		\caption{The plane mapping plot of the differential branching ratios for the decay process  $\bar B_{s} \rightarrow [\rho^{0}(\omega,\phi)\rightarrow (\pi^{+}\pi^{-})][K^{*0}\rightarrow K^{+}\pi^{-}]$.}		
		\label{fig3}
	\end{minipage}
\end{figure}
In Figs. 9 and 10, the differential branching ratios for the decay process  $\bar B_{s} \rightarrow [\phi(\rho^{0},\omega)\rightarrow K^{+}K^{-}][K^{*0}\rightarrow K^{+}\pi^{-}]$ are illustrated. For the $\bar{B_s}\to(K^+K^-)(K^+\pi^-)$ decay, integration is performed within the peak regions, with the selected ranges being ${\sqrt{s_{1}}}$ from 0.99 to 1.2 GeV and ${\sqrt{s_2}}$ from 0.85 to 0.95 $GeV$. Similarly, in Figs. 11 and 12, the differential branching ratios for the decay process $\bar B_{s} \rightarrow [\rho^{0}(\omega,\phi)\rightarrow (\pi^{+}\pi^{-})][K^{*0}\rightarrow K^{+}\pi^{-}]$ are shown. For the $\bar{B_s}\to(\pi^+\pi^-)(K^+\pi^-)$ decay, integration is also conducted within the peak regions, with the chosen ranges being ${\sqrt{s_{1}}}$ from 0.70 to 1.10 $GeV$ and ${\sqrt{s_2}}$ from 0.85 to 0.95 $GeV$.

\begin{table}[!htbh]
	\caption{ Polarization fractions $\bar B_{s} \rightarrow {(K^{+} K^{-}) (K^{+}\pi^{-})} $ and $\bar B_{s} \rightarrow {(\pi^{+} \pi^{-})( K^{+}\pi^{-})} $ decay processes.}
	\label{tab:cpvv}
	\begin{ruledtabular}  
		\setlength{\extrarowheight}{8pt}
		\setlength{\tabcolsep}{2.5mm}{ \begin{tabular}[t]{lccccc }
				Decay Channel     &Modes & $f_0(\%)$       &$f_\|(\%)$  & $f_\perp(\%)$ & ${\cal B}(10^{-6})$
				\\\hline\hline
				$\bar{B_s}\to(K^+K^-)(K^+\pi^-)$     &$\rho^{0}  K^{*0}$  &$46.65^{\pm0.32}_{\pm9.31}$ &$27.50^{\pm0.02}_{\pm2.65}$ &$25.84^{\pm0.30}_{\pm2.35}$ &$0.063^{\pm0.006}_{\pm0.039}$\\
				$$  &$\omega K^{*0}$ &$50.12^{\pm2.99}_{\pm8.66}$&$25.78^{\pm1.57}_{\pm3.84}$&$24.10^{\pm1.43}_{\pm6.82}$&$0.014^{\pm0.002}_{\pm0.015}$  \\
				$$&$\phi K^{*0}$ &$78.22^{\pm 0.53}_{\pm12.76}$   &$11.54^{\pm0.32}_{\pm3.25}$    &$10.23^{\pm0.18
				}_{\pm1.56}$     &$1.49^{\pm0.10}_{\pm0.56}$  \\
				$$ &$(\phi-\rho^{0})  K^{*0}$ &$77.43^{\pm 0.02}_{\pm4.53}$   &$11.29^{\pm0.01}_{\pm1.45}$    &$11.28^{\pm0.01}_{\pm1.16}$     &$1.51^{\pm0.07}_{\pm0.52}$  \\
				$$&$(\phi-\omega) K^{*0}$ &$76.54^{\pm0.08}_{\pm6.26}$   &$12.18^{\pm0.03}_{\pm2.01}$    &$11.28^{\pm0.05}_{\pm1.12}$     &$1.28^{\pm0.07}_{\pm0.57}$  \\
				$$&$(\rho^{0}-\omega) K^{*0}$ &$59.25^{\pm 2.36}_{\pm8.89}$   &$25.27^{\pm1.42}_{\pm5.84}$    &$15.48^{\pm0.94}_{\pm3.94}$     &$0.092^{\pm0.017}_{\pm0.043}$  \\
				$$&$(\phi-\rho^{0}-\omega) K^{*0}$ &$75.90^{\pm 0.07}_{\pm9.02}$   &$12.68^{\pm0.03}_{\pm6.57}$    &$11.41^{\pm0.05}_{\pm3.56}$     &$1.28^{\pm0.07}_{\pm0.57}$\\   \hline\hline
				
				$\bar{B_s}\to(\pi^+\pi^-)(K^+\pi^-)$     &$\rho^{0}  K^{*0}$  &$46.64^{\pm0.31}_{\pm9.19}$ &$27.51^{\pm0.01}_{\pm2.43}$ &$28.84^{\pm0.12}_{\pm2.24}$ &$0.25^{\pm0.03}_{\pm0.07}$\\
				$$ &$(\rho^{0}-\omega) K^{*0}$ &$39.65^{\pm 1.96}_{\pm12.80}$   &$36.03^{\pm1.31}_{\pm10.92}$    &$24.32^{\pm0.65}_{\pm8.54}$     &$0.38^{\pm0.11}_{\pm0.19}$  \\
				$$ &$(\rho^{0}-\phi) K^{*0}$ &$46.65^{\pm 0.32}_{\pm9.18}$   &$27.50^{\pm0.02}_{\pm2.42}$    &$25.85^{\pm0.29}_{\pm2.25}$     &$0.25^{\pm0.02}_{\pm0.06}$  \\
				&$(\rho^{0}-\omega-\phi) K^{*0}$ &$39.66^{\pm 0.023}_{\pm12.82}$   &$36.03^{\pm1.31}_{\pm10.93}$    &$24.31^{\pm0.36}_{\pm8.52}$     &$0.58^{\pm0.01}_{\pm0.11}$\\
		\end{tabular}}
	\end{ruledtabular}
\end{table}

In Table V, we present the polarization fractions and corresponding branching ratios for the four-body decays through various intermediate states. Subsequently, we analyze the central values of these results.
The branching ratio for the decay process $\bar{B_s}\to\phi(\phi\to K^+K^-) K^{*0}(K^{*0}\to K^+\pi^-)$ is two orders of magnitude higher than that for the decay processes $\bar{B_s}\to\rho^0(\rho^0\to K^+K^-) K^{*0}(K^{*0}\to K^+\pi^-)$ and $\bar{B_s}\to \omega (\omega\to K^+K^-) K^{*0}(K^{*0}\to K^+\pi^-)$. However, the polarization fractions $f_\|(\%)$ and $f_\perp(\%)$ for the decay processes $\bar{B_s}\to\rho^0(\rho^0\to K^+K^-) K^{*0}(K^{*0}\to K^+\pi^-)$ and $\bar{B_s}\to \omega (\omega\to K^+K^-) K^{*0}(K^{*0}\to K^+\pi^-)$ are approximately 2.5 times higher than those for the decay process $\bar{B_s}\to\phi(\phi\to K^+K^-) K^{*0}(K^{*0}\to K^+\pi^-)$. Due to the $\rho^{0}-\omega$ mixing, the branching ratio for the decay process $\bar{B_s}\to\rho^0(\omega)(\rho^0(\omega)\to K^+K^-) K^{*0}(K^{*0}\to K^+\pi^-)$ changes significantly compared to the direct decay $\bar{B_s}\to\phi(\phi\to K^+K^-) K^{*0}(K^{*0}\to K^+\pi^-)$ as shown in Table V. Meanwhile, the $\rho^{0}-\omega$ mixing also affects the polarization fractions: the polarization fraction $f_0$ decreases from $78.22\%$ to $59.25\%$, while $f_\|$ increases from $11.54\%$ to $25.27\%$ and $f_\perp$ increases from $10.23\%$ to $15.84\%$ when comparing $\bar{B_s}\to(\rho^0-\omega\to K^+K^-) K^{*0}(K^{*0}\to K^+\pi^-)$ with $\bar{B_s}\to\phi(\phi\to K^+K^-) K^{*0}(K^{*0}\to K^+\pi^-)$. 
 After incorporating the mixing mechanisms of $\rho^{0}-\omega$ and $\rho^{0}-\omega-\phi$, the branching ratio and the longitudinal polarization fraction $f_\|(\%)$ for the decay $\bar{B_s}\to\rho^{0}(\rho^{0}\to \pi^+\pi^-) K^{*0}(K^{*0}\to K^+\pi^-)$ increase slightly, while the transverse polarization fraction $f_0(\%)$ decreases slightly. In contrast, introducing the $\rho^{0}-\phi$ mixing mechanism has a negligible effect on both the branching ratio and the polarization fractions for this decay.

In the heavy quark limit, theoretical uncertainties arise from various sources, leading to ambiguity in the results. Specifically, the $1/m_{b}$ power correction introduces additional complexity that necessitates a thorough error analysis. In this study, the primary sources of error are as follows: First, uncertainties associated with the CKM matrix elements; Second, uncertainties stemming from hadronic interaction parameters, including form factors, decay constants, and meson wave functions.

\section{Summary and conclusion}

In this work, we introduce the $\rho-\omega-\phi$ mesons mixing mechanism and study the CP violation, regional CP violation ($A^{\Omega}_{CP}$), polarization fractions, and branching ratios in the four-body decay processes of $\bar{B}_{s} \rightarrow (K^{+} K^{-})(K^{+}\pi^{-})$ and $\bar{B}_{s} \rightarrow (\pi^{+} \pi^{-})(K^{+}\pi^{-})$ under the PQCD approach.
The interference effects arising from the mixing of $\rho$, $\omega$, and $\phi$ mesons lead to more pronounced CP violation phenomena compared to direct decay processes. ${A}^{\Omega}_{CP}$ is quantified by integrating over phase space, and distinct CP violation is observed for different mixing intermediate states when the invariant mass of the $K^{+} K^{-}(\pi^{+} \pi^{-})$ pair is within specific ranges.
When the invariant masses of $K^{+}K^{-}$ pair and $\pi^{+}\pi^{-}$ pair are in the resonance region, the CP violation can be detected experimentally by reconstructing $\rho$, $\omega$ and $\phi$ mesons. This study may provide some reference for the future detection and research of the LHCb experiment.

In 2010, the Large Hadron Collider (LHC) at CERN successfully carried out proton-proton collisions at $7$ TeV. The LHC is designed with a center-of-mass energy of $14$ TeV and a luminosity of $ L = 10^{34} cm^{-2} s^{-1}$. The production cross-section of $b\bar b$
at the LHC is anticipated to be $0.5 mb$, yielding approximately $0.5 \times 10^{12}$
bottom events annually \cite{QuarkoniumWorkingGroup:2004kpm}. The LHCb experiment, leveraging these events, focuses on studying CP violation and rare decays within the $b$ hadron system. During Runs 1 and 2, LHCb amassed substantial data, while ATLAS and CMS gathered $5 fb^{-1}$, $20 fb^{-1}$, and $150 fb^{-14}$ at $\sqrt{s}= 7$ TeV, $8$ TeV, and $13$ TeV, respectively.
The High-Luminosity Large Hadron Collider (HL-LHC) and its potential upgrade to the High-Energy Large Hadron Collider (HE-LHC), a $27$ TeV proton-proton collider, are poised to further advance flavor physics research. The experimental sensitivity of the HL-LHC is expected to markedly improve, with ATLAS and CMS planning to record $3000 fb^{-1} $ of data and LHCb's Upgrade II aiming to capture $300 fb^{-1}$ \cite{Cerri:2018ypt}. These datasets will offer new opportunities for the study of CP violation phenomena.

 \section*{Acknowledgements}
This work was supported by  Natural Science Foundation of Henan (Project no.232300420115), and National  Science Foundation of China (Project no.12275024).

\end{spacing}
\end{document}